\documentclass[journal]{IEEEtran}
\usepackage[utf8]{inputenc}
\usepackage{xcolor,soul,framed} 
\usepackage{multicol}
\usepackage{booktabs}
\usepackage{tabularx}

\colorlet{shadecolor}{yellow}
\usepackage[pdftex]{graphicx}
\graphicspath{{../pdf/}{../jpeg/}}
\DeclareGraphicsExtensions{.pdf,.jpeg,.png}

\usepackage[cmex10]{amsmath}
\usepackage{array}
\usepackage{mdwmath}
\usepackage{eqparbox}
\usepackage{url}
\newcommand*\rot{\rotatebox{90}}

\usepackage{url}
\usepackage{comment}
\usepackage{ifthen}
\usepackage{amssymb}
\usepackage{balance}
\usepackage[normalem]{ulem} 
\usepackage{xcolor}
\usepackage{array}
\usepackage{booktabs}
\usepackage{graphicx}
\usepackage{xspace}
\usepackage{supertabular}
\usepackage[breakable,skins]{tcolorbox}

\usepackage{enumitem}

\usepackage{tikz}

\newtcolorbox{activitybox}[1][]{%
	breakable,
	enhanced,
	colback=blue!5,
	colframe=blue!40!black,
	coltitle=white,
	#1
}
\newtcolorbox{findingbox}[1][]{
		breakable,
		enhanced,
		colback=white, 
		arc=0.8em,
	#1
}

\newboolean{showcomments}
\setboolean{showcomments}{true} 
\ifthenelse{\boolean{showcomments}}
{\newcommand{\nb}[2]{
		\fcolorbox{black}{yellow}{\bfseries\sffamily\scriptsize#1}
		{\sf\small$\blacktriangleright$\textit{#2}$\blacktriangleleft$}
	}
	
}
{\newcommand{\nb}[2]{}
	
}

\usepackage{booktabs}
\usepackage{array}
\usepackage{colortbl}
\usepackage{multirow}
\usepackage{longtable}

\newcolumntype{v}[1]{>{\raggedright \hspace {0pt}}p{#1}}
\newcolumntype{G}[1]{>{\columncolor{gray90}}#1}

\definecolor{Gray}{gray}{0.8}
\definecolor{gray25}{gray}{0.25}
\definecolor{gray50}{gray}{0.50}
\definecolor{gray75}{gray}{0.75}
\definecolor{gray90}{gray}{0.9}

\begin{document}

\bstctlcite{IEEEexample:BSTcontrol}
    \title{How do Practitioners Perceive the Relevance of Requirements Engineering Research?}
    
  \author{Xavier Franch, 
  Daniel Mendez,~\IEEEmembership{Member,~IEEE}, 
  Andreas Vogelsang,~\IEEEmembership{Member,~IEEE}, 
  Rogardt~Heldal,~\IEEEmembership{Member,~IEEE},
  Eric Knauss,~\IEEEmembership{Member,~IEEE},
  Marc Oriol, 
  Guilherme H. Travassos, 
  Jeffrey~C.~Carver,~\IEEEmembership{Senior Member,~IEEE}, 
  Thomas Zimmermann,~\IEEEmembership{Senior Member,~IEEE}

  \thanks{X. Franch is with Universitat Polit\`ecnica de Catalunya, Barcelona (e-mail: franch@essi.upc.edu).}
  \thanks{D. Mendez is with Blekinge Institute of Technology, Sweden, and with fortiss GmbH, Germany (e-mail: Daniel.Mendez@bth.se).}
  \thanks{A. Vogelsang is with University of Cologne, Germany (e-mail: vogelsang@cs.uni-koeln.de).}
  \thanks{R. Heldal is with Western Norway University of Applied Sciences (e-mail: rohe@hvl.no).}
  \thanks{E. Knauss  is with University of Gothenburg (e-mail: eric.knauss@cse.gu.se).}
  \thanks{M. Oriol is with Universitat Polit\`ecnica de Catalunya, Barcelona (e-mail: moriol@essi.upc.edu).}
  \thanks{G.H. Travassos is with Federal University of Rio de Janeiro, Brazil (e-mail: ght@cos.ufjr.br).}
  \thanks{J.C. Carver is with the University of Alabama (e-mail: carver@cs.ua.edu).}
  \thanks{T. Zimmermann is with Microsoft Research, Redmond, USA (e-mail: tzimmer@microsoft.com).}

}

\markboth{IEEE TRANSACTIONS ON SOFTWARE ENGINEERING, VOL.~XX, NO.~YY, MARCH~2020
}{Xavier \MakeLowercase{\textit{et al.}}: RE-Pract}

\IEEEtitleabstractindextext{%
\begin{abstract}
\emph{Context}: The relevance of Requirements Engineering (RE) research to practitioners is vital for a long-term dissemination of research results to everyday practice. Some authors have speculated about a mismatch between research and practice in the RE discipline. However, there is not much evidence to support or refute this perception. 
\emph{Objective}: This paper presents the results of a study aimed at gathering evidence from practitioners about their perception of the relevance of RE research and at understanding the factors that influence that perception. 
\emph{Method}: We conducted a questionnaire-based survey of industry practitioners with expertise in RE. The participants rated the perceived relevance of 435 scientific papers presented at five top RE-related conferences. 
\emph{Results}: The 153 participants provided a total of 2,164 ratings. The practitioners rated RE research as essential or worthwhile in a majority of cases. However, the percentage of non-positive ratings is still higher than we would like. Among the factors that affect the perception of relevance are the research's links to industry, the research method used, and respondents' roles. The reasons for positive perceptions were primarily related to the relevance of the problem and the soundness of the solution, while the causes for negative perceptions were more varied. The respondents also provided suggestions for future research, including topics researchers have studied for decades, like elicitation or requirement quality criteria. 
\emph{Conclusions}: The study is valuable for both researchers and practitioners. Researchers can use the reasons respondents gave for positive and negative perceptions and the suggested research topics to help make their research more appealing to practitioners and thus more prone to industry adoption. Practitioners can benefit from the overall view of contemporary RE research by learning about research topics that they may not be familiar with, and compare their perception with those of their colleagues to self-assess their positioning towards more academic research. 
\end{abstract}

}

\maketitle
\IEEEdisplaynontitleabstractindextext

\section{Introduction}

Software engineering (SE) research, in particular applied SE research, has a unique symbiotic relationship between theoretical and practical knowledge. 
Hence, in SE, either considered as a whole or in its specific activities, we need an ongoing discussion about the alignment between the theoretical contributions of researchers and practical problems.

While researchers have recently evaluated this alignment within SE research as a whole~\cite{Lo15} and empirical SE research~\cite{Carver16}, there is no such evaluation within requirements engineering (RE) research. 
This knowledge gap is particularly concerning because of the doubts about the practical relevance of RE over the years. 
These doubts are fueled, for example, by the uniqueness of some types of systems, like open-source-based~\cite{Alspaugh13} or commercial-off-the-shelf-based~\cite{Franch04}, the pervasiveness of agile development approaches, and the increasing popularity of data-driven decision making\footnote{ \url{https://janbosch.com/blog/index.php/2017/04/30/the-end-of-requirements/}}. 
For instance, agile approaches encourage practitioners to reduce irrelevant documentation and improve knowledge management practices~\cite{Hummel2013}, which are frequently at odds with more traditional RE~\cite{Wohlrab2019a}. 
Given their biases towards the importance of the topics they have studied for years, it is often very difficult for researchers to judge the practical relevance of their own research.

Therefore, this paper addresses the gap through an empirical study. 
Using a questionnaire-based survey, we gathered practitioners' perception of the relevance of the RE research embodied in a sample of 435 RE papers published in major conferences.
Based on 2,164 ratings from 153 respondents, we answer the following questions:
\begin{itemize}
    \item \emph{How do practitioners perceive the relevance of RE research?} We present an overview of the perceived relevance and discuss how the relevance differs depending on the role of the practitioner or the age of the paper. 
    \item \emph{What contextual factors influence this perception?} We investigate the impact of paper types, research methods, the affiliations of authors, and general topics on the perceived relevance.
    \item \emph{What guidance can be given for maximizing the relevance of future RE research?} We provide the reasons given by practitioners for their ratings and an analysis of the topics that they suggest for future research. 
\end{itemize}

In addition to exploring the relevance of RE research, we reflect on factors that influence practitioners' perception of relevance. The results provide a practical perspective on why practitioners value some research topics and insight into the type of research on which the RE community should, and should not, focus. 
This type of insight can guide researchers who are new to RE or who are interested in changing their research to topics that are more relevant to industry. 
The results provide a list of potential research topics, many of which are well known but have not received adequate attention, suggesting a lack of academia--industry collaboration.

\section{Background}
\label{sec:RelWork}

The practical relevance of SE research has been an object of attention since the 90s~\cite{Potts93}~\cite{Glass94}. 
For instance, Glass echoed some reports affirming that ``research was ignoring practice almost entirely and, in addition, ignoring the notion of practical application"~\cite{Glass94}. 
The problem of the practical relevance of SE research is still a matter of debate, not just in general, but also in specific domains, such as global SE~\cite{Beecham14}, and specific activities such as software architecture~\cite{Malavolta13}, testing~\cite{Garousi17} and RE, which is our focus. 

Despite their active research and promising contributions to the field, RE researchers still want to understand how their solutions apply to practical problems~\cite{Ameller2020, Mendez17}. 
Several authors have noted a mismatch between research and practice resulting in a gap between researchers and practitioners~\cite{Ghaisas14}.
This mismatch has been present for at least 20 years both in SE and in other topics like information systems engineering (e.g., motivating a panel at CAiSE, the flagship conference in the area~\cite{KaindlMyl00}).  
This discussion begun a fruitful and relevant research agenda~\cite{Kaindl02} that continued through discussions at scientific conferences related to RE~\cite{Mahaux13}. 
The set of concerns raised by these discussions suggest a variety of questions for the SE community in general and some of its sub-disciplines in particular.

An initial study into the practical relevance of SE research investigated how Microsoft practitioners perceived the research published in ICSE, ESEC/FSE, and FSE papers between 2009 and 2014~\cite{Lo15}. 
A replication of that study investigated how practitioners from multiple companies perceived the relevance of the research published in 156 ESEM papers between 2011 and 2015~\cite{Carver16}.
The results of the ESEM study showed: (1) papers with industrial authors did not have a higher perceived relevance than papers without an industrial author and (2) whether the paper summary was written by the paper's original authors or by the researchers made little difference in its perceived relevance.
To build upon the call to broaden the scope of their study to include other topics, this paper picks up that thread for RE. 

In RE, our own prior work begun to address this issue by proposing a protocol to answer the following questions: (1) Do practitioners perceive academic RE research to be relevant to their work and why (or why not) and (2) How can scholars make RE research (even more) relevant to practitioners?~\cite{Franch2017}
Building upon that research protocol, this paper reports the results of our study of the relevance of RE research to practitioners’ based upon practitioners' perceptions of the relevance of RE research described in research papers.
Our work here presents the in-depth results of the study. 
We previously presented a very brief overview of selected results~\cite{VMF19}, with the goal of raising awareness among practitioners and receiving early feedback from the research community. 
The current paper extends these preliminary results by providing extensive analysis of contextual factors, qualitative data, and broad interpretation of the results. 

\section{Study Design}
\label{sec:Design}
This section overviews the study design.
Our replication package, which is an open dataset and open materials, provides the complete protocol, the datasets, and the analysis instrumentation~\cite{dataset}.

\subsection{Research Objective, Questions and Approach}
\label{sec:RQs}
The goal of this study is \textit{to assess the practical relevance of RE research as perceived by practitioners}. 
To address this goal, we developed the research questions in Table~\ref{tab:rqs}. 
\begin{table}[!htb]
    \centering
    \caption{Research questions of the study}
    \label{tab:rqs}
    \begin{tabularx}{\columnwidth}{@{}lX@{}}
    \toprule
    ID & Text \\
    \midrule
    RQ 1 & How do practitioners perceive the relevance of RE research? \\[1ex]
    RQ 2 & Which properties of RE research influence perceptions? \\[1ex]
    RQ 3 & What reasons do practitioners give to justify their perceptions of relevance? \\[1ex]
    \hspace{1em}RQ 3.1 & What reasons do practitioners give to justify their positive perceptions of relevance? \\
    \hspace{1em}RQ 3.2 & What reasons do practitioners give to justify their negative perceptions of relevance?\\[1ex]
    RQ 4 & Which research problems should the RE research community address according to practitioners?\\
    \bottomrule
    \end{tabularx}
\end{table}

Evaluating the relevance of RE research to practitioners would require the consideration of every RE research effort, which is not feasible. 
Therefore, we chose scientific publications as representative of this research. 
We argue that this approach, adopted by the antecedent studies cited above~\cite{Lo15, Carver16}, is suitable because the scientific community regularly disseminates its mature results through scientific publications. 
Instead of building our study on scientific papers, we could have used other approaches, for instance analysing practitioner perceptions of RE technology effectively transferred to industry through grey literature or even dedicated interviews. 
Still, the set of all scientific publications related to RE is too large and we needed to select a subset, as detailed in Section~\ref{sec:design-dataset}. 
This subset was presented to practitioners who provided ratings through a survey as explained in Section \ref{sec:instrument}.

RQ1 focuses on the practitioners' overall perceptions of the relevance of RE research and any factors which might affect their perceptions, such as their project role or year in which the research was published. 
This question provides the basis for more detailed analyses about how practitioners rate the specific research topics in the papers. 

RQ2 focuses on the extent to which perceived relevance depends upon properties of the publications, including the topic addressed, the research methods used, the authors' ties to industry, and the publication venue.

The remaining research questions gather opinions from practitioners to justify their perception of relevance. 
In RQ3, we are particularly interested in understanding the reasons for the ratings given, both positive ratings (RQ3.1) and negative ratings (RQ3.2).
In RQ4, we identify situations practitioners encounter in their daily work that could be good candidates for researchers to investigate. 
These two questions provide researchers with useful feedback for planning future research. 

\subsection{Team}
We composed a team of researchers that meet the following criteria: 
\begin{itemize}
    \item have different backgrounds and experiences to ensure the study design and execution are richer and less prone to bias; 
    \item allow for distribution of tasks in the data preparation and collection, e.g. emerging from the paper summarization, while increasing the validity by having the ability to cross-check the analysis results among the team;
    \item transfer authors’ experiences from similar studies on relevance to practitioners~\cite{Carver16, Lo15}, and on RE-related topics, including NaPiRE~\cite{Mendez17} and NFR4MDD~\cite{Ameller2020}; and 
    \item gain access to a wider network of practitioners. 
\end{itemize}

Following previous work on authors’ attribution and contribution~\cite{Brand15}, Table~\ref{tab:authorshipdetails} summarizes the roles of the team members. 
We provide additional details in the prior publication~\cite{Franch2017}. 

\begin{table}[!htb]
\scriptsize
\centering
\caption{Authorship details: Roles and contribution (alphabetical order). 'S' stands for support or discussion role.}
\label{tab:authorshipdetails}
\begin{tabularx}{\columnwidth}{@{}Xcccccccccc@{}}
\toprule
\textbf{Contribution}  & \rot{\textbf{J. Carver}} & \rot{\textbf{X.\ Franch}} & \rot{\textbf{R.\ Heldal}} & \rot{\textbf{E.\ Knaus}} & \rot{\textbf{D.\ Mendez}} & \rot{\textbf{M.\ Oriol}} & \rot{\textbf{G.H.\ Travassos}} & \rot{\textbf{A.\ Vogelsang}} & \rot{\textbf{T.\ Zimmermann}}\\ \hline
Conceptualisation &  & X &  &  & X &  &  &  &  \\
Administration &  & X &  &  & X &  &  & X &  \\
Instrument Design &  & X &  &  & X &  &  &  &  \\
Instrument Review & X & X & X & X & X & X & X & X & X \\
Paper Summaries & X & X & X & X & X & X & X & X &  \\
Validation of Summaries & X & X &  & X & X &  & X & X &  \\
Survey Implementation &  &  &  & & X &  & &  &  \\
Survey Validation & X & X &  & X &  &  & X &  &  \\
Scripts Development &  &  &  &  &  &  & & X &  \\
Coding &  & X &  & & X &  & & X &  \\
Quantitative Analysis &  & X & X & X & X & S & & X &  \\
Qualitative Analysis  &  & X &  & & S & & & S &  \\
Writing - Draft & S & X & S & S & X &  & S & X &  \\
Writing - Review \& Editing & X & X & X & X & X & X & X & X & X \\
\bottomrule
\end{tabularx}
\end{table}

\subsection{Dataset}
\label{sec:design-dataset}

Driven by our research approach (Section~\ref{sec:RQs}), we had to make three decisions related to the dataset.

First, we had to choose the publication venues. 
Aligning with the decision of prior work~\cite{Lo15}, we considered major RE-related conferences and purposefully excluded journals (see Section~\ref{sec:ExternalValidity} for a discussion).
We included: 1) the two flagship RE-specific academic research conferences: IEEE RE\footnote{Throughout the paper, we use ``IEEE RE'' instead of the conference acronym ``RE'' to avoid any ambiguity with our use of ``RE'' as an abbreviation for ``Requirements Engineering'' as a discipline.} and REFSQ; 2) three general top-ranked SE conferences where RE papers commonly appear: ICSE, ESEC\slash FSE, and ESEM. 
We considered all papers from IEEE RE and REFSQ for inclusion/exclusion (see below).
For the general SE conferences, we first filtered out non-RE papers.

Second, we had to define the inclusion/exclusion criteria. 
We chose to include: 1) full research papers (excluding all short papers like tool demos and vision papers); 2) full industry papers (i.e. published in an industry track). We further limited the study to papers published between 2010 and 2016 to focus on the most recent papers when launching our survey in 2017. 
Table~\ref{tab:dataset} lists the final number of papers from each venue, the number of industry papers, and the number of ratings received (as detailed in Section \ref{sec:responses}). 
Note that for our study, we define ``industry papers'' to include not only papers published in industry tracks but also papers with at least one industry author (more details in Section~\ref{sec:rq2}).

\begin{table}[!htb]
    \centering
    \caption{Paper per venue}
    \label{tab:dataset}
    \begin{tabular}{@{}lrrrr@{}}
    \toprule
    Venue & Papers & Industry Papers & Ratings (all) & Ratings (industry)\\
    \midrule
    IEEE RE     & 245 & 92 (37.6\%) & 1237 &  452 (36.5\%) \\
    REFSQ     & 113 & 28 (24.8\%) & 540 & 125 (23.1\%)\\
    ICSE     & 39 & 7 (17.9\%) & 195 & 31 (15.9\%) \\
    ESEM     & 22 & 9 (40.9\%) & 121 & 51 (42.1\%)\\
    ESEC\slash FSE; FSE     & 16 & 3 (18.8\%) & 71 & 17 (23.9\%)\\
    Total     & 435 & 139 (32.0\%) & 2164 & 676 (31.2\%)\\
    \bottomrule 
    \end{tabular}
\end{table}

Third, we had to decide what information to present to practitioners to make their evaluation of relevance.
We believed that requiring respondents to read full papers would have resulted in very few (if any) responses. 
Therefore, we used paper summaries, following the example of previous studies~\cite{Lo15, Carver16}. 
While we could have used the paper abstracts as the summary, our analysis of the abstracts identified too much variation in quality, style, and length.
A previous study showed these types of variations could affect the perception of relevance~\cite{Carver16}. 
Therefore, to ensure consistency, we decided to write the summaries ourselves, using the following template :

\begin{quote}
\textit{
\textbf{A} [type of work] \textbf{for} [purpose of the work] \textbf{in order to} [expected benefit from the work from a practical point of view]
}
\end{quote}

For example: 
\begin{quote}
\textit{
``An experiment with students for comparing two requirements elicitation techniques when instantiating software product lines (SPL) in order to understand which approach is more suitable for eliciting requirements in SPL'' [P196]\footnote{Throughout this manuscript we refer to research papers in the form ``[P{\it xxx}]'', where {\it xxx} denotes the unique identifier of the publication.}.
}
\end{quote}
All information on the included publications is available in our replication package~\cite{dataset}. 

While we attempted to use the template for all summaries, we did not rigidly enforce use of the template because for some papers (e.g., for literature reviews) it was difficult, if not impossible, to use it. 
In addition, because some papers did not explicitly describe the paper's intention (i.e. the text after \textbf{in order to}), we did not speculate and used only the authors’ words, when available. 
Despite these potential impediments, we found the use of the template helpful for ensuring similar descriptions for all papers and providing a homogeneous dataset.

\subsection{Instrument}
\label{sec:instrument}
We conducted the survey using an online questionnaire (available in our replication package~\cite{dataset}) to enable us to reach a wide community of practitioners across the globe. 
In addition, because the survey asked for critical feedback on research, we made the questionnaire anonymous.

In a quest for simplicity, we designed the questionnaire with three primary questions:
\begin{enumerate}
    \item First, the questionnaire presented the respondent with 15 randomly chosen paper summaries. 
    The respondents provided their opinion on the importance of the research described in each summary using a 4-point rating scale: \textit{Unwise}, \textit{Unimportant}, \textit{Worthwhile}, or \textit{Essential}. This asymmetric survey response scale is inspired by the Kano model~\cite{Kano84} and used by previous similar studies~\cite{Lo15, Carver16} making it easier to compare results. Respondents could leave a summary unrated if they could not judge the relevance of the research or if they did not have an opinion. 
    The responses to this question help answer RQ1 and RQ2.
    \item Then, the questionnaire presented the respondent with one of their highest-ranked summaries and one of their lowest-ranked summaries and asked them to provide a brief explanation for the ratings. The responses to this question are used to answer RQ3. 
    \item Finally, the questionnaire asked the respondents to describe problems they faced in their work that could benefit from research. 
    The responses to this question are used to answer RQ4.
\end{enumerate}
The questionnaire also gathered the following demographics: 
\begin{itemize}
    \item \textit{Position}: main role and level of engagement in RE duties (responsible, contributor, user of requirements, or none).
    \item \textit{Background}: working experience (in years) and academic degree (in computer science-related field or not).
    \item \textit{Context}: industry sector, size of the team, and country.
    \item \textit{Projects}: type of software typically developed (e.g., embedded, information systems). 
\end{itemize}
The questionnaire concluded with a free text field where respondents could provide any further comments. 
However, this question did not provide any useful information..

\subsection{Participants}
We used the following channels to recruit participants:
\begin{itemize}
    \item \textit{RE Community e-mail Lists}: REonline\footnote{re-online@it.uts.edu.au}, contacts coming from previous studies such as NaPiRE\footnote{\url{www.napire.org}}~\cite{Mendez17}, and regional lists such as dist-jisbd\footnote{distjisbd@lcc.uma.es} (Spain).
    \item \textit{The IREB organization}\footnote{\url{www.ireb.org}}: list of certified RE professionals.
    \item \textit{Authors’ networks}: explicit invitations to industrial partners from across the world, either as participants or to help with dissemination of the survey in their networks.
    \item \textit{General Dissemination}: via social media such as Twitter.
\end{itemize}

\subsection{Data analysis}

We compute several statistics to characterize the overall perception practitioners have about RE research. 
Following the measures used in previous studies~\cite{Begel14, Lo15}, who justified their choice according to Kitchenham and Pfleeger's recommendations~\cite{Kitchenham08} (see~\cite{Begel14} for details), we calculate the proportion of \textit{Essential} (best response), \textit{Essential or Worthwhile} (positive feedback), and \textit{Unwise} (worst response) ratings. These metrics recognise \textit{Unimportant} responses as neutral responses that do not require the same level of analysis as the others. 

More formally, let \textit{E}, \textit{W}, \textit{Ui}, and \textit{Uw} denote the number of Essential, Worthwhile, Unimportant, and Unwise ratings received: 
\begin{itemize}
    \item \textit{E-score}: The percentage of ratings that are Essential: $\operatorname{\mathit{E-score}} = E/(E+W+\mathit{Ui}+\mathit{Uw)}$
    \item \textit{EW-score}: The percentage of ratings that are Essential or Worthwhile: $\operatorname{\mathit{EW-score}} = (E + W) / (E+W+\mathit{Ui}+\mathit{Uw})$
    \item \textit{U-Score}: The percentage of ratings that are Unwise: $\operatorname{\mathit{U-score}} = \mathit{Uw} /(E+W+\mathit{Ui}+\mathit{Uw})$
\end{itemize}
We then can compute these statistics for different groups of respondents or summaries, e.g., all ratings, ratings by certain demographics, ratings for specific conferences, or ratings for category of papers.

\section{Coding}
This section describes our coding process for the natural language responses.
We had to perform coding to classify research topics (Section~\ref{sec:coding-topics}) and to categorize the free text responses (Section~\ref{sec:coding-opentext}). 

\subsection{Classification of Research Topics}
\label{sec:coding-topics}
Our focus was to analyze the topics described in the summaries and their underlying papers.
For a couple of reasons, our goal was not to analyze the perceived relevance of individual research contributions. 
First, from a methodological point of view, the number of ratings for each summary is rather small (median of five). 
Second, our interest is to identify overall perception and trending topics rather than specific research topics. 
Therefore, we assess the relevance of research topics by aggregating ratings from multiple summaries about similar topics. 

Instead of working top-down from an existing body of knowledge (e.g., SWEBOK), we worked bottom-up from the respondents' answers. 
Therefore, we applied inductive coding to create a list of RE topics well-suited for the responses gathered in the study (Table~\ref{tab:topic-categories}). 
The list defines four top-level categories, of which two have subcategories. 
Each category contains a list of topics. Though coding was inductive, we aimed at using consolidated knowledge in RE. For instance, in the case of Requirements Quality, we referred mostly to characteristics mentioned in the ISO/IEC/IEEE 29148 standard~\cite{ISO29148} and added only a couple of additional codes mentioned frequently in the responses, e.g. uncertainty, defined as the degree to which requirements change during the development process.
We associated each summary with one or more topics (similar to tags).

\begin{table*}
    \centering
    \caption{Research topic categories}
    \label{tab:topic-categories}
    \begin{tabular}{@{}p{2cm}ll l lll l@{}}
    \toprule
    \multicolumn{3}{c}{Challenge} & 
    Documentation &
    \multicolumn{3}{c}{Subject matter} & 
    Phase \\
    \cmidrule{1-3}\cmidrule{5-7}
    Requirements quality & 
    People & 
    Process & &
    First level & 
    Second level & 
    Third level & 
    \\
    \midrule
    general & collaboration & automation & general & goals & & & elicitation \\
    completeness & communication & decision making & business model & functional & & & documentation \\
    consistency & skills & formalization & goal model & quality $\Rightarrow$ & generic & & V \& V \\
    feasibility & subjectivity & prioritization & feature model & architecture & reliability & & negotiation \\
    traceability & creativity & standardization & state machine & tests & performance \&  & & management \\
    unambiguousness & other & visualization & UML diagram & process & \hspace{1em}efficiency & & \\
    understandability & & reuse & natural language & feature & usability & & \\
    necessity & & quality assurance & use case & risk & security & & \\
    verifiability & & execution & user story & other & compatibility & & \\
    uncertainty & & evolution & scenarios & & maintainability & & \\
    other & & modeling & prototype & & portability& safety & \\
    & & configuration & rules & & other \hspace{1em} $\Rightarrow$ & dependability & \\
    & & other & review report & & & sustainability & \\
    & &  & other & & & privacy & \\
    & &  &  & & & legal \&  regulations & \\
    \bottomrule
    \end{tabular}
\end{table*}

To create the list of topics and associate them with the list of summaries, the three qualitative analysis authors (Table~\ref{tab:authorshipdetails} -- referred to as \textit{QL-authors} in the rest of the section) performed the following three-step process. 

\textbf{Step 1}: The QL-authors created an initial topic list based on their experience and background acquired during the preparation phase (e.g., the writing of the summaries). 
Using that list, each QL-author tagged ten randomly selected summaries (all authors used the same ten summaries). 
Then, the QL-authors compared their results, discussed deviations, extended the list of topics, and discussed open questions about the tagging process. 
After this first round of coding, the QL-authors identified ten new topics, renamed seven topics, and moved two topics from one category to another. 
In addition, the QL-authors decided that: (i) they would use only the content of the summaries for the tagging process (not the full paper) and (ii) a summary could have multiple tags. 

\textbf{Step 2}: The QL-authors repeated step one on a new set of ten randomly selected summaries. 
This step resulted in three new topics. 
At this point, the QL-authors reached a consensus and decided to proceed with coding the entire dataset. 

\textbf{Step 3}: Each QL-author tagged all 435 summaries with topics from finalized topic list.
Instead of discussing all summaries after the rating process, they applied a resolution scheme to determine the final tagging of topics. 
The underlying principle was when at least two QL-authors agreed on a tag, the tag remained.
Table~\ref{tab:res-scheme} shows the definition of the seven rules that form the resolution schema applied to the topic tags for each summary.
For example, rule C2 prevents the assignment of topic \textit{topic X} to a summary with \textit{topic X} if only one QL-author tagged the summary with that topic. 
Rule C5 tags a summary with \textit{topic X} if two QL-authors tagged the summary with that topic but the third disagreed.
The last column of the table shows the percentage of cases for which each rule applied. 
Over 90\% of decisions used rules C1, C2, C6, C7, in which at least two QL-authors agreed.

\begin{table}
    \centering
    \caption{Resolution Scheme}
    \label{tab:res-scheme}
    \begin{tabular}{@{}cccccr@{}}
    \toprule
    Rule & Rater 1 & Rater 2 & Rater 3 & Decision & Applied\\
    \midrule
    C1   & none     & none  & none  &   none    & 57.4\%  \\    
    C2   & X        & none  & none  &   none    & 11.1\%  \\
    C3   & X        & Y     & none  &   none    & 2.2\%  \\
    C4   & X        & Y     & Z     &   none    & 1.1\%  \\
    C5   & X        & X     & Y     &   X       & 5.4\%  \\
    C6   & X        & X     & none  &   X       & 6.2\%  \\
    C7   & X        & X     & X     &   X       & 16.7\%  \\
    \bottomrule
    \end{tabular}
\end{table}

\subsection{Coding of Free Text}
\label{sec:coding-opentext}
We needed to code the reasons for positive and negative ratings (RQ3) and the suggestions for research topics (RQ4). To classify the free text responses about the reasons for positive and negative ratings, we used content analysis with inductive category formation~\cite{Mayring00}. 
In the case of research topics, we were able to reuse some of the topics included in the taxonomy presented in Table~\ref{tab:topic-categories} (see Section V-F for details). 

To perform the coding, two QL-authors independently conducted a preliminary analysis and held a consolidation meeting to produce a draft set of categorized codes. 
They then used these codes to individually re-analyze the responses and held another consolidation meeting to produce a new coding, if necessary. 
While the categories themselves did not change, the authors added, removed, or changed codes for some answers. 
As the last step, the third QL-author made a final check and proposed a few minor suggestions, which the three QL-authors discussed and agreed upon.

During the coding process, the QL-authors removed some responses for the following reasons: 1) they were empty, 2) they were not understandable, 3) they did not match the question (e.g., a positive reason when the question asked the reason for a negative answer), 4) the respondent did not understand the summary (based either on the respondent's explicit statement or on the content of the response), or 5) the respondent stated the summary was too abridged to answer the question. 
In addition, they translated a small number of responses from German to English.

The codes that emerged from this process are not mutually exclusive. 
Hence, a good number of answers had more than one code attached, even in the same category. 
Sections~\ref{sec:Results-RQ3} and \ref{sec:Results-RQ4} contain the final list of codes for the qualitative answers.

\section{Results}
We structure the results around the RQs. 
Then, Section~\ref{sec:analysis} provides analysis and discussion of these results.

\subsection{Overview of Responses}
\label{sec:responses}
We received responses from 154 practitioners. 
Because we relied on open mailing lists, we cannot calculate the response rate. 
The respondents provided a total of 2,164 ratings. 
A large share of the respondents (73.4\%, 113 individuals) rated all 15 summaries presented to them. 
Only 4.6\% (7 respondents) rated less than 13 summaries. 
The lowest number of ratings given by a participant was 5.

For each summary, we computed the consensus, i.e. the level of agreement, based on the formula by Tastle and Wierman~\cite{Tastle07}, depicted in equation~\eqref{eq:1}.

\begin{equation}
\textit{Consensus}(X)= 1 + \sum_{i=1}^{n} p_i \log_2 \left ( 1 - \frac{\left | X_i - \mu_X  \right |}{ d_X} \right )
\label{eq:1}
\end{equation}

where:
\begin{itemize}
\item $n$ is the number of rating categories. In our case, $n=4$ (i.e. Unwise, Unimportant, Worthwhile, Essential).
\item $p_i$ is the percentage of respondents who considered the summary to be in the $i$th rating category.
\item $X_i$ is the numeric value of the $i$th rating category. In our case, from 1 to 4.
\item $\mu_X$ is the mean of the ratings given by the respondents for the summary.
\item $d_X$ is the width of $X$, that is, $X_{max} - X_{min}$. In our case, $d_X$ = 3.
\end{itemize}

\begin{figure}
    \centering
    \includegraphics[width=\columnwidth]{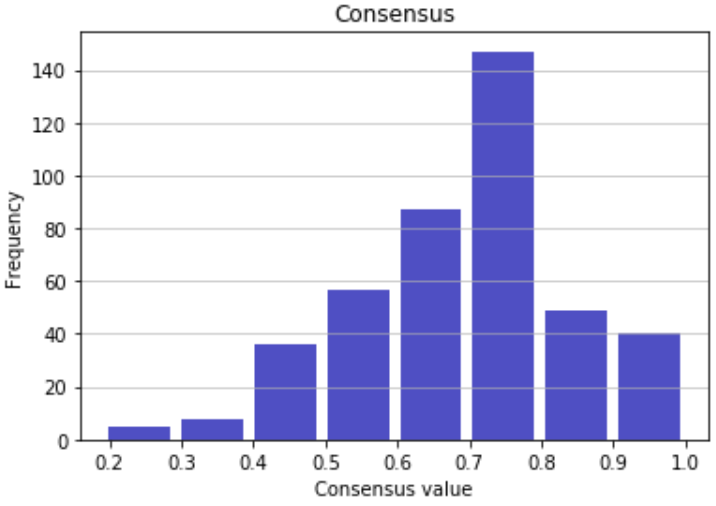}
    \caption{Consensus between respondents regarding the ratings}
    \label{fig:consensus}
\end{figure}

This process led to 429 consensus values (one per rated summary\footnote{Only six papers did not receive any ratings and were excluded from the analysis.}) ranging from 0 (complete disagreement) to 1 (complete agreement). 
Fig.~\ref{fig:consensus} depicts the distribution of these values. 
The average consensus among respondents’ ratings was 0.69. 
For example, a summary receiving one \textit{unimportant}, five \textit{worthwhile}, and four \textit{essential} ratings has a consensus value of 0.69. 
To determine whether the number of ratings is correlated with the consensus value, we calculated the Spearman rank correlation (because both values are not normally distributed). 
There is a statistically significant correlation (p-value < 0.001), but the correlation is weak (rho is -0.16). 
In other words, the number of ratings has little, if any, influence on the consensus value.
The source data for the calculations can also be found in our replication package~\cite{dataset}.  

\subsection{Demographics}
\label{sec:demographics}

\textit{Country.} 
We received responses from 32 countries, including all continents, ranging from 93 (Europe) to 2 (Africa). 
We received the largest number of responses from Germany (39; 25.3\%), followed by the United States (14; 9.1\%), and Brazil and China (10; 7.8\%). 

\textit{Experience and education.} 
Respondents have varied levels of experience with an average of more than 11.5 years (although some respondents declared not to be sure about the total time or responded using modifiers as ``approximately'' or ``more than''). 
For education, 116 respondents (76.0\%) have a degree in Computer Science or related field.

\textit{Role and involvement.} 
The largest group of respondents (64; 41.6\%) usually assume the role of Requirements Engineer or Business Analyst. 
The other roles are more distributed, with a good number (30; 19.5\%) working in roles not in the given list, such as consultant, marketing person, or systems engineer. 
Independent of their role, 101 respondents (65.6\%) stated they were responsible for RE tasks, 83 (53.9\%) actively contributed to RE tasks, and 58 were consumers of requirements (37.7\%). 
The number of respondents in each level includes the number of respondents in the previous level.

\textit{Context.} 
Respondents work in teams of all sizes.
The largest group of respondents (68; 44.2\%) work in medium-sized teams (5 to 10 members). 
Respondents work on the following types of systems: embedded systems (32; 20.8\%), information systems (72; 46.8\%), or both (37; 24.0\%), with a few (12; 7.8\%) working on other types of systems like apps or desktop tools. 
Respondents also worked in diverse domains, the most popular being: automotive (24; 15.6\%), telecommunications (17; 11.0\%), and banking (9; 5.8\%).

\subsection{RQ 1: How do practitioners perceive the relevance of RE research?}
\label{sec:Results-RQ1}
\begin{figure}
    \centering
    \includegraphics[width=\columnwidth]{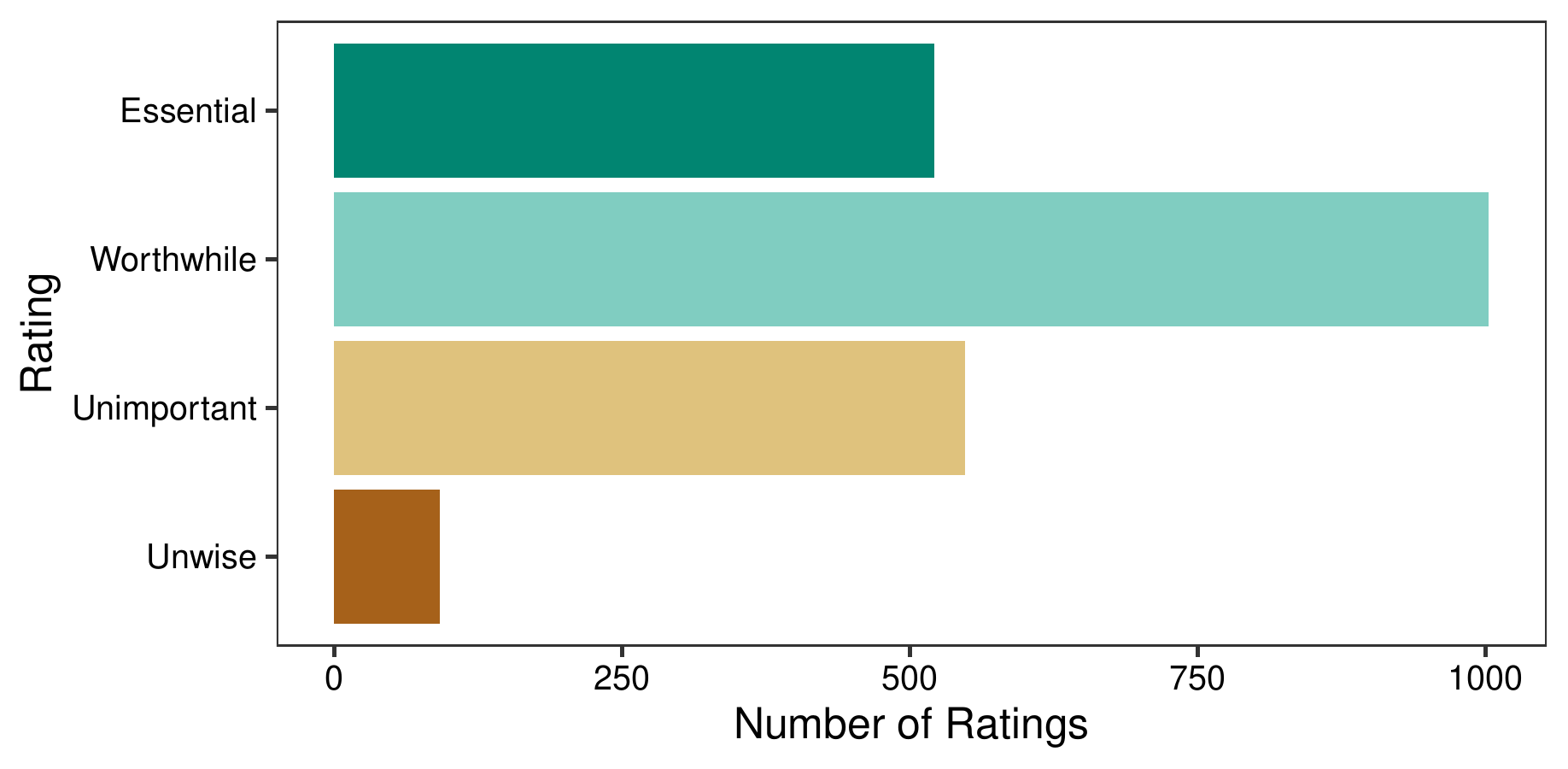}
    \caption{Overall Perception of RE research relevance.}
    \label{fig:OverallPerception}
\end{figure}

Figure~\ref{fig:OverallPerception} shows the overall perception ratings for all paper summaries. 
A majority of practitioners consider RE research positively (EW-score: 0.70) or even essential (E-score: 0.24). 
However, in 30\% of the cases, respondents rated research negatively, and in almost 5\% even as unwise (U-score: 0.04). 

If we focus on papers rather ratings, we find 304 papers (69.9\%) with more positive than negative ratings. 
Of those papers, 106 (24.4\%) had only positive ratings, compared to 17 (3.9\%) that received only negative ratings. 
The average number of ratings for papers with only positive ratings is 8.1, which is much higher than the number of ratings for papers with only negative ratings (1.6).

Focusing on the respondents, their overall perception was positive, with 130 (84.4\%) providing more positive than negative ratings, including 21 (13,6\%) who provided only positive ratings.
Only 2 respondents provided only negative ratings.

\begin{figure}
    \centering
    \includegraphics[width=\columnwidth]{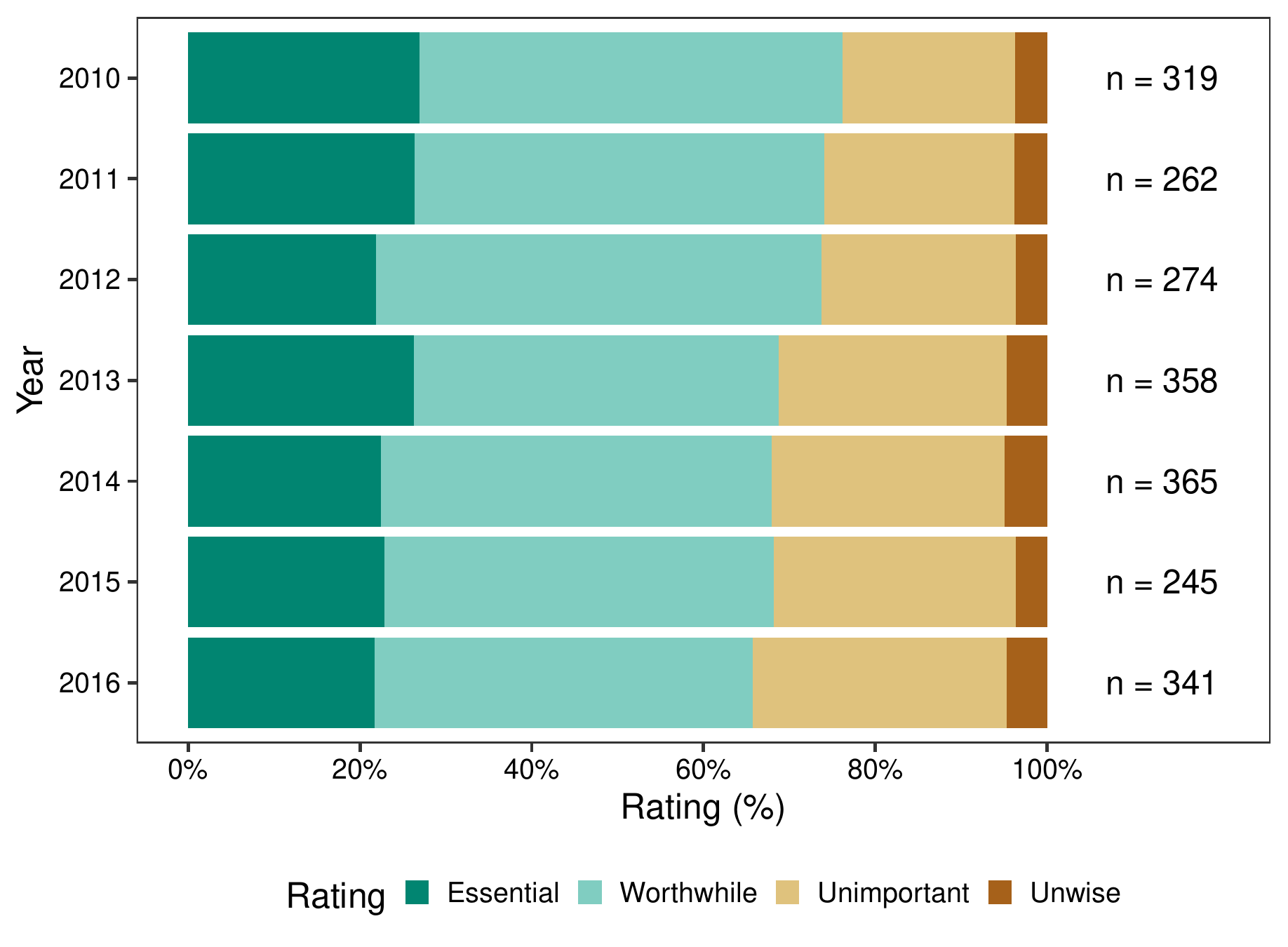}
    \caption{Perception of RE research relevance measured along the years of the respective publications.}
    \label{fig:PerceptionYearOfPublication}
\end{figure}

From a temporal perspective, Figure~\ref{fig:PerceptionYearOfPublication} shows a slight but consistent downward trend in perceived relevance for papers published in more recent years (EW-Score of 0.76 in 2010 declines to 0.65 in 2016). 

\begin{figure}
    \centering
    \includegraphics[width=\columnwidth]{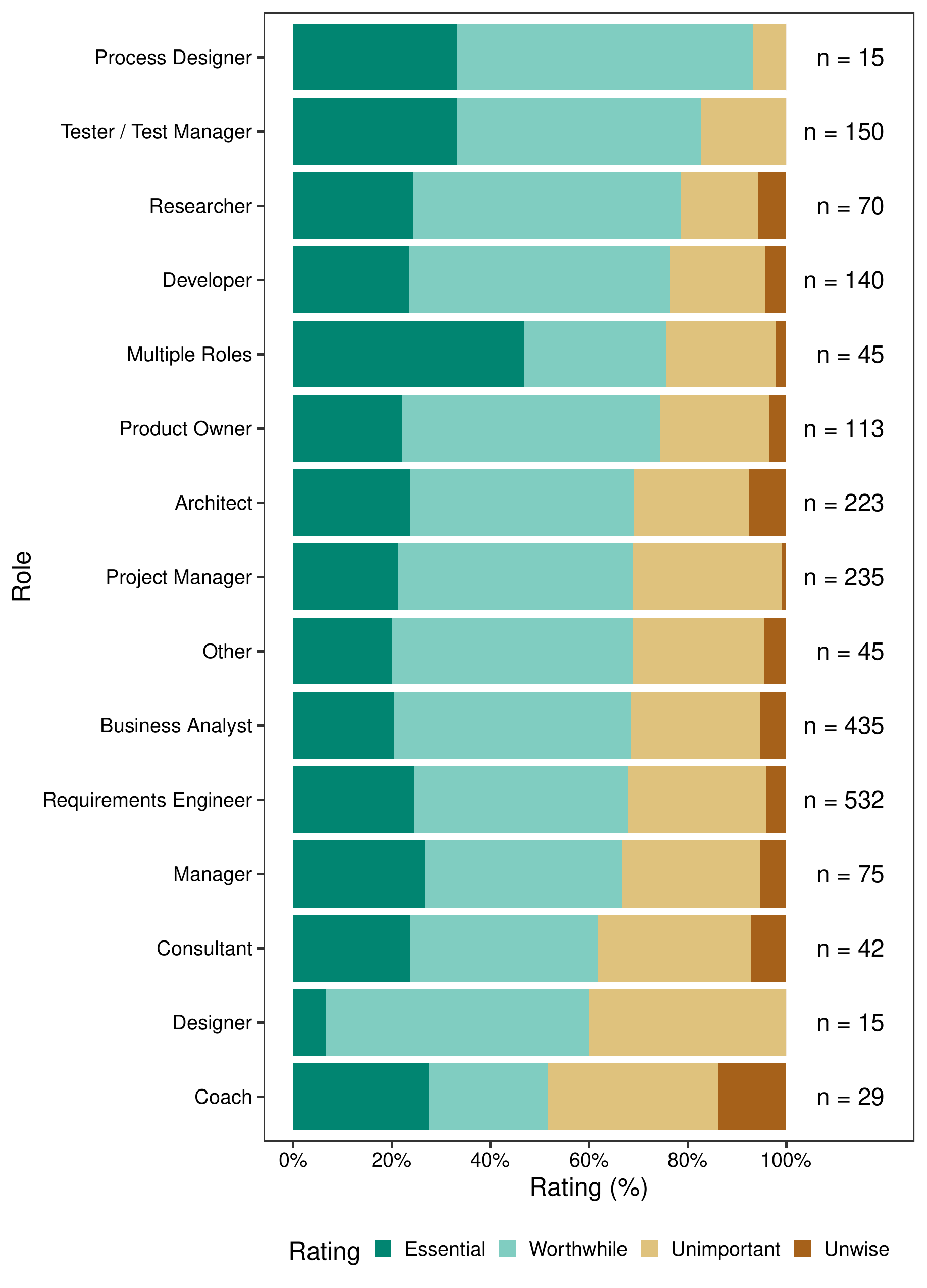}    
    \caption{Rating of relevance according to the respondents' roles. 
    }
    \label{fig:PerceptionRole}
\end{figure}

Figure~\ref{fig:PerceptionRole} summarizes the perceived relevance of RE research for respondents in different roles. 
While the results are diverse, they are overall positive.
Only respondents working as \textit{coaches} rated the summaries slightly more negative.
However, this result needs to be taken with caution given the low number of coaches in our sample (only 29 ratings). 

\subsection{RQ 2: Which properties of RE research influence perceptions?}
\label{sec:rq2}
To address this question, we analyze the effects of four characteristics on perceived relevance: (1) publication venue, (2) ties to industry, (3) research method, and (4) paper topics.

\subsubsection{Does the perception of the relevance of a paper depend on the publication venue of the paper?}

\begin{figure}
    \centering
    \includegraphics[width=\columnwidth]{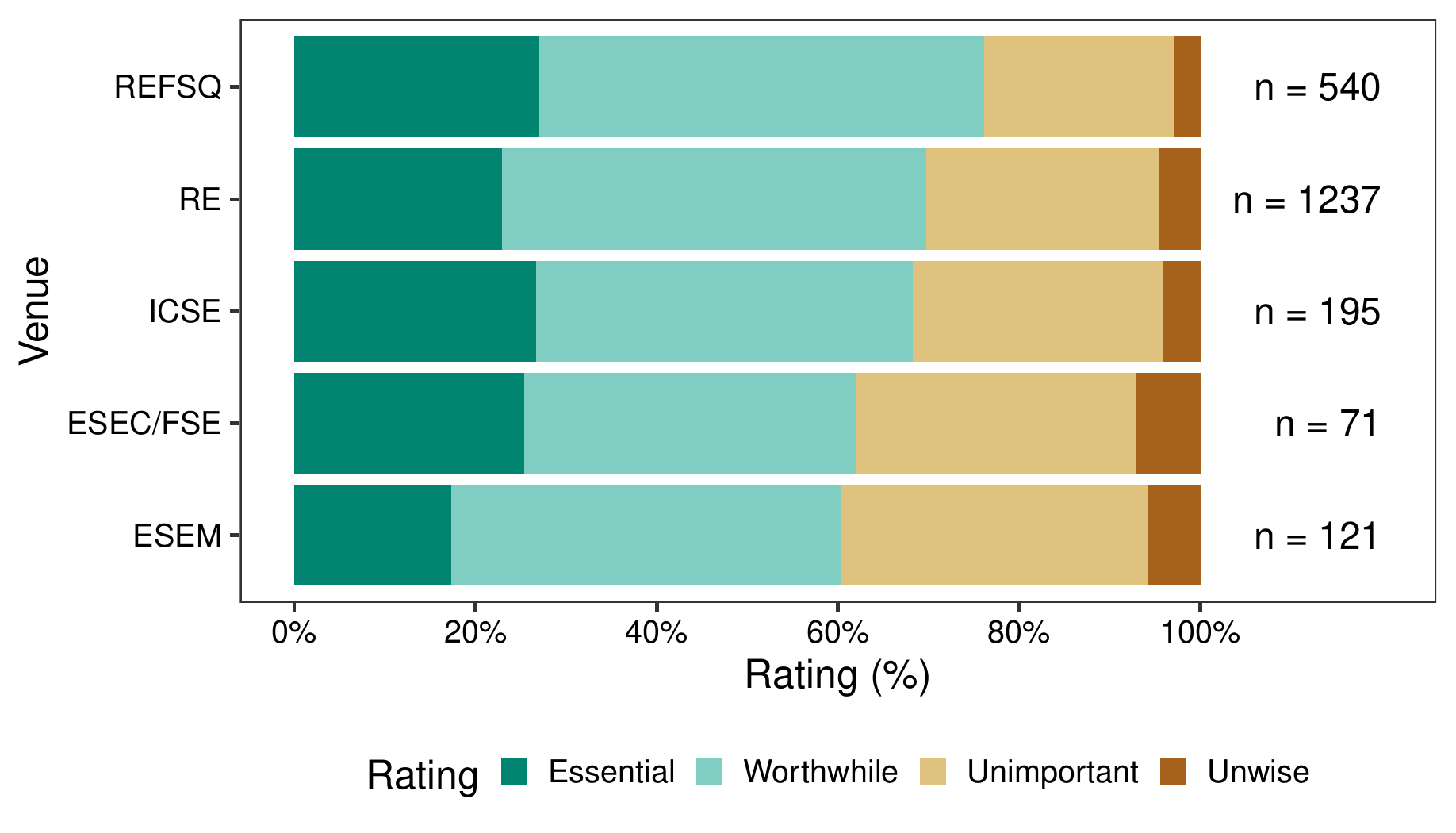}
    \caption{Perception of RE research relevance about the venue.}
    \label{fig:PerceptionVenue}
\end{figure}

Figure~\ref{fig:PerceptionVenue} shows a slight tendency towards more positive ratings for papers in venues specifically focused on RE (REFSQ and IEEE RE; combined EW-Score: 0.71) than for papers in more general venues (combined EW-score: 0.64).

\subsubsection{Does the perception of the relevance of a paper depend on the kind of institution, industrial or academic, the paper's authors work in?}

\begin{figure}
    \centering
    \includegraphics[width=\columnwidth]{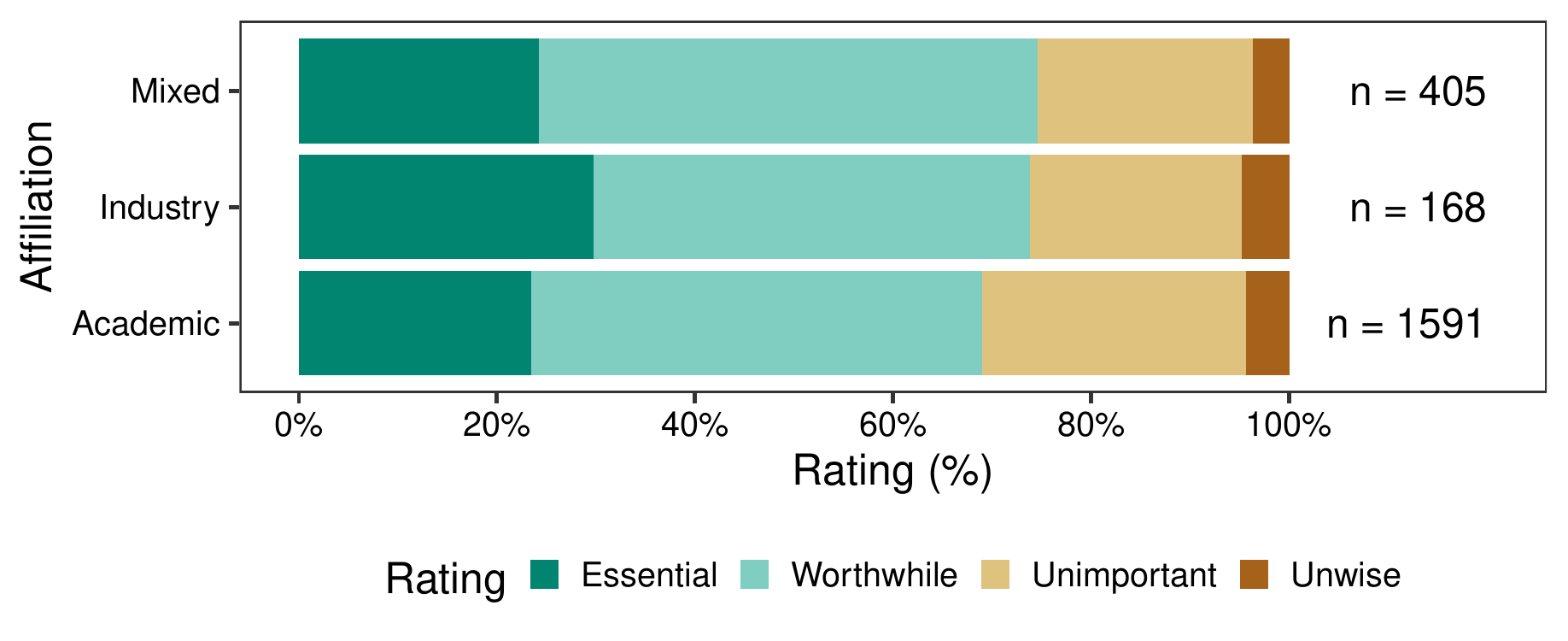}
    \caption{Rating grouped by authors' affiliation.}
    \label{fig:PerceptionAuthorAffiliation}
\end{figure}

We grouped the overall ratings according to the affiliations of the authors. 
We distinguish between papers with only academic authors (i.e. universities, research institutions, or transfer institutes such as Fraunhofer), papers with only industrial authors (i.e., companies), and papers with mixed author teams. 
The results in Figure~\ref{fig:PerceptionAuthorAffiliation} show little difference among the groups. 
Respondents' perception of relevance is slightly higher for papers with industry participation (pure industry or mixed). 
Papers with only industry authors achieved the highest ratio of \emph{Essential} ratings. 

\begin{figure}
    \centering
    \includegraphics[width=\columnwidth]{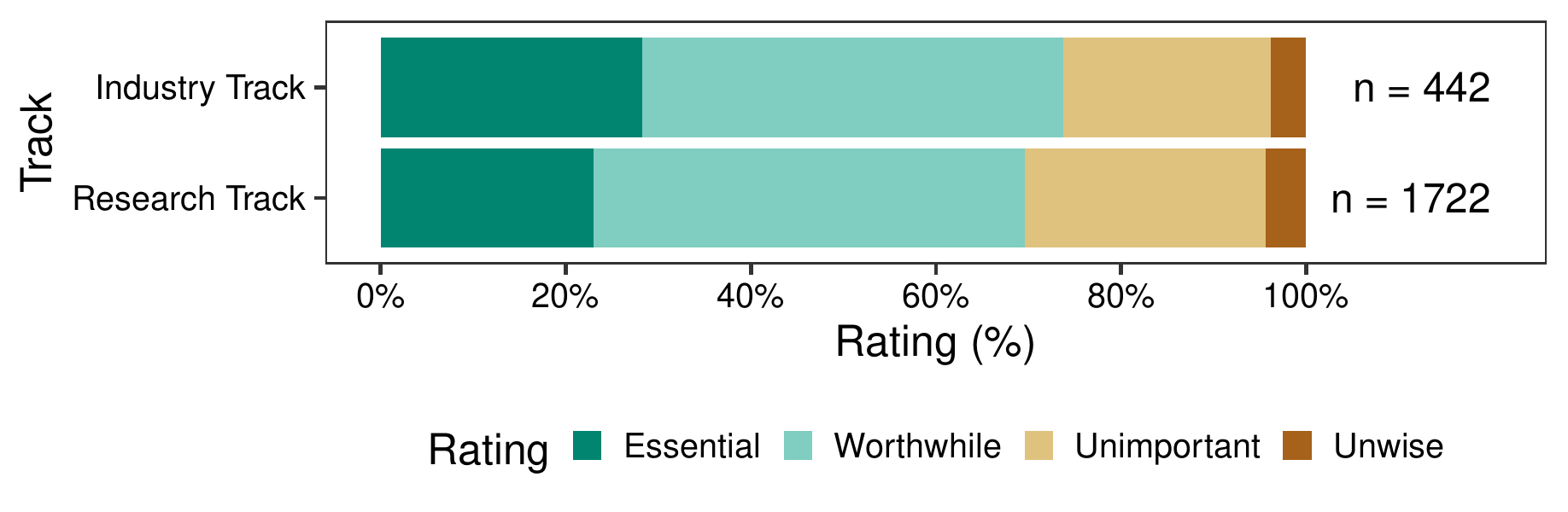}
    \caption{Ratings grouped by submission track}
    \label{fig:RatingsTrack}
\end{figure}

In addition, Figure~\ref{fig:RatingsTrack} shows the distribution of ratings depending upon whether the paper appeared in an industry track or a research track.
The results show that respondents perceive papers in an industry track as slightly more relevant than papers in a research track. 

\subsubsection{Does the perception of the relevance of a paper depend on the research method used to do the paper's research?}

\begin{figure}
    \centering
    \includegraphics[width=\columnwidth]{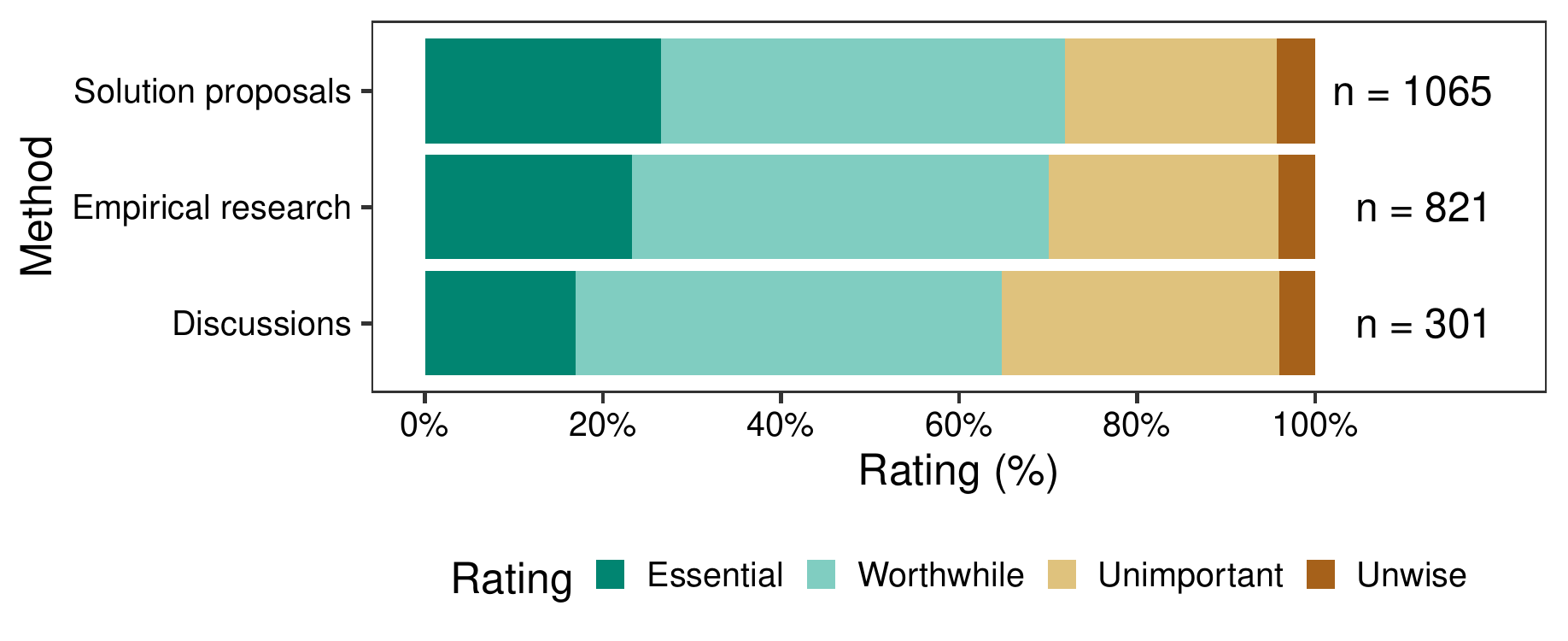}
    \caption{Overall ratings in dependency to the research method}
    \label{fig:OveralRatingsType}
\end{figure}

Figure~\ref{fig:OveralRatingsType} shows how the perception of relevance varies across papers of the following types:

\begin{enumerate}
    \item  ``Solution proposal'' papers contain technical contributions in the form of methodologies, reference models, or technologies.
    \item ``Empirical research'' papers analyze --from an empirical perspective-- contemporary real-world phenomena via interrogations (e.g., interviews), observational studies (e.g., case studies), or intervention studies (e.g., controlled experiments).
    \item ``Discussion'' papers provide philosophical elaborations, reviews (e.g., literature reviews), opinions\slash position statements, or experience reports.
\end{enumerate}

The EW-scores show a slight preference for solution proposal papers over empirical research papers, with discussion papers receiving the lowest score. 

\begin{figure}
    \centering
    \includegraphics[width=\columnwidth]{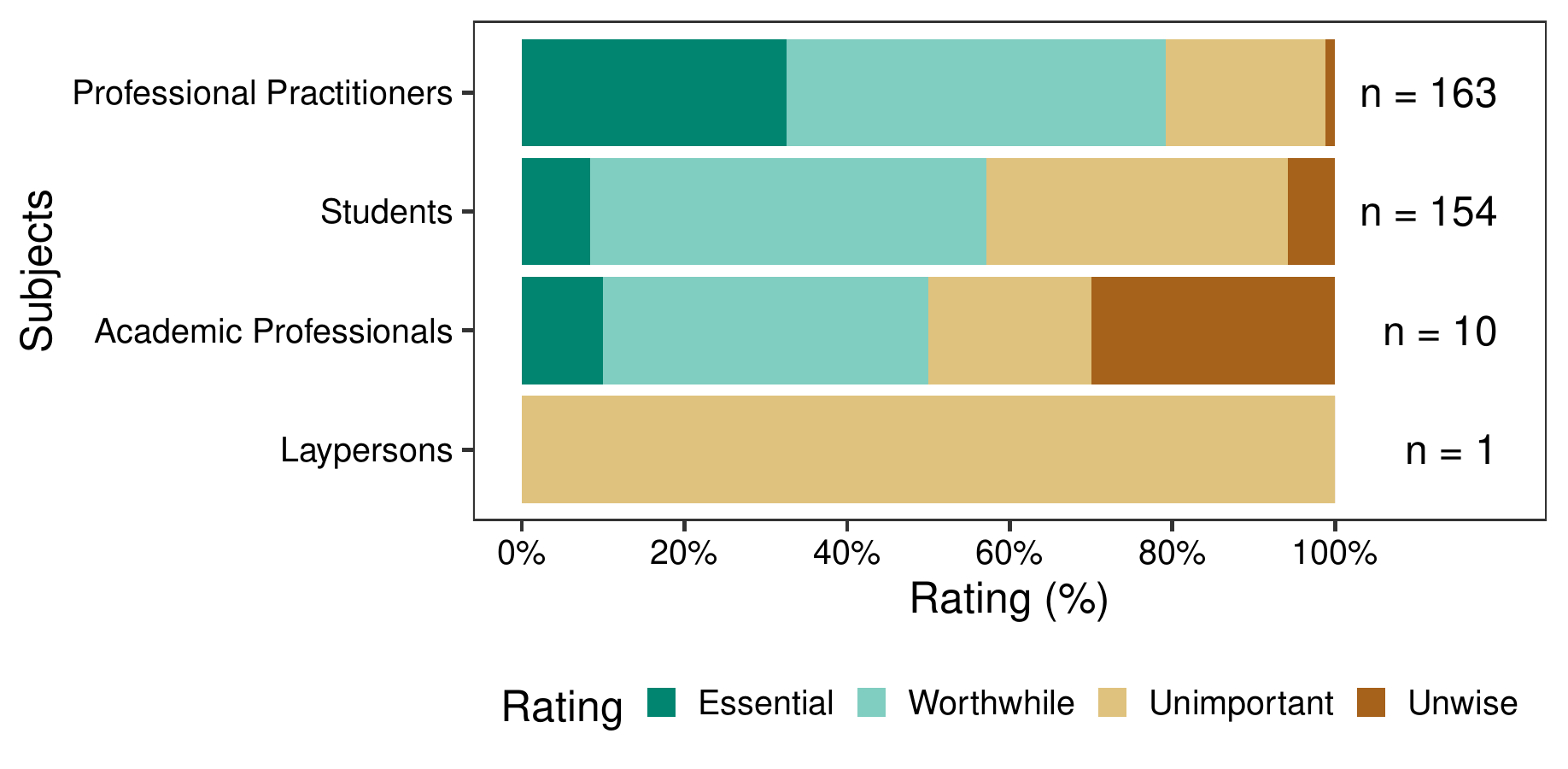}
    \caption{Overall ratings in dependency to involved subjects.}
    \label{fig:OverallRatingsSubjects}
\end{figure}

A clearer picture emerges when we consider the type of study participants (Figure~\ref{fig:OverallRatingsSubjects}). 
Not surprisingly, respondents perceived papers with professional practitioners as participants to be the most relevant.
They perceived papers with students and academic professionals as participants to be similar in terms of EW-score. 
Interestingly, a large share of the respondents considered papers with academic professionals as participants to be unwise (U-Score: 0.3; please note that the number of associated ratings is rather low).

\subsubsection{Does the perception of the relevance of a paper depend on the paper's research topics?"}

For reasons of conciseness and simplicity, we summarize only the five highest-rated and five lowest-rated research topics (Tables~\ref{tab:HIGLYRATEDTOPICS} and \ref{tab:CRITICALLYRATED}, respectively). 
We exclude research topics of type ``other'' from these tables because they lack specificity.
In addition, we only included research topics that had more than one paper and more than five ratings associated with it. 
The full data is available in our online dataset~\cite{dataset}. 

Table~\ref{tab:HIGLYRATEDTOPICS} shows the five most highly-rated research topics sorted in terms of their E-score (descending) followed by their EW-scores (descending).
For these topics, more than 30\% of the ratings judged the topic as essential. 
The topics span several categories including: requirements quality, people aspects, and specific documentation styles.

\begin{table*}
\footnotesize
    \centering
        \caption{Highly rated research topics}
    \label{tab:HIGLYRATEDTOPICS}
    \begin{tabular}{@{}llrrrrr@{}}
\toprule
Topic & Category & Nr.\ of papers & Nr.\ of ratings & E-score & EW-score & U-score \tabularnewline
\midrule
General requirements quality & Quality & 8 & 37 & 0.43 & 0.89 & 0.00
\tabularnewline
Subjectivity & People & 4 & 19 & 0.42 & 0.73 & 0.05
\tabularnewline
Scenarios & Documentation & 5 & 26 & 0.38 & 0.80 & 0.03
\tabularnewline
Collaboration & People & 15 & 83 & 0.33 & 0.75 & 0.04
\tabularnewline
User stories & Documentation & 4 & 18 & 0.33 & 0.68 & 0.00
\tabularnewline
\bottomrule
\end{tabular}

\end{table*}

Table~\ref{tab:CRITICALLYRATED} highlights the five most critically rated research topics sorted in terms of their U-score (descending). 
Among these topics, between 8\% and 11\% of the ratings were highly critical (i.e., rated the related research as unwise). 
The topics span several categories including: requirements quality, people aspects, and specific documentation styles. 
While for some topics, such as goal models and creativity, the high U-score matches a low E and EW-scores, there are also cases where a high U-score contrasted with a rather high EW score (e.g., the topics of use cases or necessity). 

\begin{table*}
    \centering
        \caption{Critically rated research topics}
    \label{tab:CRITICALLYRATED}
    \begin{tabular}{@{}llrrrrr@{}}
\toprule
Topic & Category & Nr.\ of papers & Nr.\ of ratings & E-score & EW-score & U-score \tabularnewline
\midrule
Use cases & Documentation & 10 & 53 & 0.24 & 0.73 & 0.11
\tabularnewline
Negotiation & Phase & 4 & 19 & 0.15 & 0.78 & 0.10
\tabularnewline
Necessity & Quality & 3 & 10 & 0.20 & 0.80 & 0.10
\tabularnewline
Creativity & People & 7 & 44 & 0.15 & 0.63 & 0.09
\tabularnewline
Goal model & Documentation & 11 & 56 & 0.07 & 0.64 & 0.08
\tabularnewline
\bottomrule
    \end{tabular}

\end{table*}

\subsection{RQ3. What reasons do practitioners give for justifying their perceptions of relevance?}
\label{sec:Results-RQ3}
To answer RQ3, we analyze the free-text answers respondents gave to justify their most positive and most negative ratings.

\subsubsection{RQ 3.1 What reasons do practitioners give for justifying their positives perception of relevance?}

After removing three invalid responses, we analyzed the justifications for positive ratings from 119 out of the 154 respondents (77.3\%), which produced 204 code instances. 
Table~\ref{tab:PositiveCodes} presents codes, grouped by categories, resulting from the process described in Section~\ref{sec:coding-opentext}. 
Overall, respondents gave justifications related primarily to \textbf{Problem Relevance} of the problem and the \textbf{Utility} of the solution. 
Therefore, we created a category for each of these justifications.
We identified two other categories of codes: (i) \textbf{Impact} - the possible consequences of the research presented and (ii) \textbf{Source} - the respondent’s experience.

\begin{table*}
    \centering
    \caption{Codes for positive ratings}
    \label{tab:PositiveCodes}
    \begin{tabularx}{0.9\textwidth}{@{}lXr@{}}
    \toprule
    Code & Definition & Occurrences\\
    \midrule
    \textbf{Problem relevance} & the respondent justifies his positive rating in terms of the relevance of the problem & \textbf{93}\\
    \hspace{1em}general & the problem is of interest per se, without reference to the relevance to industry & 40\\
    \hspace{1em}industry & the problem is of interest for practitioners & 31\\
    \hspace{1em}contextual & the problem is of interest for some area, development method, etc.& 9\\
    \hspace{1em}organizational & the problem is of interest for the organization of the respondent & 8\\
    \hspace{1em}personal & the problem is of personal interest for the individual respondent & 5\\ [1ex]
    \textbf{Utility} & the respondent justifies his positive rating in terms of the utility of the solution& \textbf{89}\\
    \hspace{1em}adequate & the respondent is confident that the solution presented to the problem follows the right approach& 30\\        \hspace{1em}plausible & the respondent considers that the solution presented to the problem may work (but s/he is not sure; this is the difference with the former) & 17\\ 
    \hspace{1em}unsatisfactory & the respondent considers that existing solutions are not satisfactory for some reason (not efficient, inaccurate, \dots)& 15\\
    \hspace{1em}not existing & the respondent is not aware of any solution for the problem & 12\\
    \hspace{1em}complex & the respondent considers that any solution to the problem is intrinsically difficult& 10\\
    \hspace{1em}scarce & the respondent reports that there are very few existing solutions to the problem & 5\\ [1ex]    \textbf{Impact}& the respondent justifies his positive rating in terms of the possible consequences of the research & \textbf{16}\\
    \hspace{1em}on individual & the respondent informs that the research reported would benefit himself (e.g., knowledge, skills, \dots)&10\\    \hspace{1em}on organization & the respondent argues that the research reported would benefit the dynamics of the organization & 6\\ [1ex]
    \textbf{Source} & the respondent informs about the source of his rating& \textbf{6}\\
    \hspace{1em}experience & the respondent sustains his positive rating because of his experience & 6\\
    \bottomrule
    \end{tabularx}
\end{table*}

The \textbf{Problem relevance} and \textbf{Utility} categories had a similar number of code instances.
However, the number of respondents who provided those code instances leans slightly towards \textbf{Problem relevance} with 86 respondents to only 77 for the code instances in \textbf{Utility}.

An interesting follow-up result is that 49 respondents provided justifications from both categories, e.g., ``Ambiguities are a critical source of issues. An automatic double check could help'' [P9;R110]\footnote{We refer to respondents in the form “[Ryyy]”, where yyy is a unique identifier. When the respondent provides an opinion on a paper, the reference becomes “[Pxxx;Ryyy], where xxx is a unique identifier. 
All information on papers, respondents, and responses is available in our online dataset~\cite{dataset}.}  

In terms of the number of instances, three codes dominated. 
First, 40 code instances (19.6\% of the total, issued by 33.6\% of respondents) justified the positive ratings in terms of their general relevance to RE without further details, e.g., ``non-functional reqs have the biggest impact on architecture and may not be left aside'' [P22;R28]. 
Next, 31 code instances (15.2\%; 26.1\%) referred also to relevance but in relation to industry, e.g., ``Companies and businesses still use a high amount of textual documents which can be used for RE purposes'' [P309;R76]. 
Third, 30 code instances (14.7\%; 25.2\%) reflect the perception that the solution followed the right approach, e.g., ``Terminology is essential for the interpretation of requirements'' [P215;R35]. 
Other common codes correspond to the following situations (see Table~\ref{tab:PositiveCodes} for details):

\begin{itemize}
\item The proposed solution seems plausible (``Correlating understanding of non-functional requirements to project success would be of interest'' [P128;R74]).
\item Existing solutions are unsatisfactory (``Establishing and maintaining trace-links is still cumbersome'' [P2;R116]).
\item Unaware of the existence of any solution for the addressed problem (``If a method could help teams with little knowledge and experience in dealing with security, it would be extremely valuable'' [P100;R57]).
\item Existing solutions are intrinsically complex (``To translate legal requirements into IT requirements needs a lot of translating effort between specialists'' [P224;R26]).
\item The proposal is of interest for individuals (``Understanding the applicability of commonly used techniques in different contexts is always helpful as it sharpens the own perception'' [P44;R102]).
\end{itemize}

\subsubsection{RQ 3.2 What reasons do practitioners give for justifying their negative perception of relevance?}

Of the 116 responses to this question, we had to remove 19 that were not valid, resulting in 97 respondents (63.0\%). Table~\ref{tab:NegativeCodes} presents 17 emerging codes grouped into five categories. Some respondents disagreed with a significant aspect of the research, while others were more concerned about the actual industrial need or the difficulty of making the research actionable. 
In addition, some respondents did not trust the proposal because they were not convinced about some aspect related to its quality. 
Last, a few respondents had concerns about whether the research was actually about RE or about whether the research would have industrial impact.

\begin{table*}
    \centering
    \caption{Codes for negative ratings}
    \label{tab:NegativeCodes}
    \begin{tabularx}{0.9\textwidth}{@{}lXr@{}}
    \toprule
    Code & Definition & Occurrences\\
    \midrule
    \textbf{Unnecessary} & the respondent justifies the negative rating by the low impact of the work&\textbf{35}\\
    \hspace{1em}not needed - universal & the respondent thinks that the research is not of interest for the industry in general&15\\
    \hspace{1em}not needed - contextual & the respondent thinks that the research does not solve an industrial need in his/her specific context&12\\
    \hspace{1em}not interesting & the respondent does not question the work but simply considers it as not interesting&6\\
    \hspace{1em}old fashioned & the respondent thinks that the work is not essential for industry anymore (could have been in the past)&2\\[1ex]
    \textbf{Difficult to apply} & the respondent justifies the negative rating by the difficulty to apply the work&\textbf{31}\\ 
    \hspace{1em}too specific & the respondent thinks the work can only be applied under very specific conditions&11\\
    \hspace{1em}not realistic & the respondent questions the practical applicability of the approach, possibly pointing out some aspect disregarded in the paper&8\\
    \hspace{1em}too complicated & the respondent thinks the approach is too complicated to apply in reality (e.g., diversity of contexts)&6\\    \hspace{1em}not efficient & the respondent questions the efficiency of the approach (even if the approach looks sound)&5\\
    \hspace{1em}too subjective & the respondent thinks that the research is too subjective&1\\[1ex] 
    \textbf{Disagreement} & the respondent justifies the negative rating by disagreeing with the work&\textbf{26}\\
    \hspace{1em}not agreed & the respondent disagrees with the research approach or the  research hypothesis presented in the paper (strong disagreement)&18\\
    \hspace{1em}not convinced & the respondent is not convinced by the research approach or the research hypothesis presented in the paper (weak disagreement)&8\\[1ex]
    \textbf{Low quality} & the respondent justifies the negative rating by the poor quality of the work& \textbf{13}\\
    \hspace{1em}weak evidence & the respondent is not convinced by the data and\slash or population that supports the research (quantity, procedure, profile, \dots)&5\\
    \hspace{1em}not state of the art & the respondent thinks that other approaches exist addressing the same problem&4\\
    \hspace{1em}too vague & the respondent understands the approach but misses a more thorough description in order to understand its applicability.&3\\
    \hspace{1em}not solid & the respondent thinks the research lacks substance (questions the solidity of the work)&1\\[1ex] 
    \textbf{Overall critique} & the respondent justifies the negative rating by providing a general argument&\textbf{7}\\
    \hspace{1em}general objection & the respondent does not question the particular research but poses an objection that is more related to the very nature of research&4\\
    \hspace{1em}not RE & the respondent thinks that the research is not about RE but another topic&3\\
    \bottomrule
    \end{tabularx}
\end{table*}

The most common negative justification is lack of necessity, followed closely by difficulty of applying and disagreement.
Contrary to the positive justifications, only a few responses produced multiple codes (14 or 14.3\%; in all the cases only two codes). 

In terms of individual codes, the four codes mentioned most often were (see Table~\ref{tab:NegativeCodes} for details):
\begin{itemize}
    \item The respondent is in total disagreement with the research. Usually, the response provides some explanation (``The decision of which feature next is not based on uncertainty but an economic or technical risk view'' [P305;R95]).
    \item The proposal is unnecessary in practice (``The industry has never go too far from natural language specifications'' [P214;R132]).
    \item The proposal is unnecessary in a particular context, usually the respondent’s organization or even the respondent as an individual (``The easiest answer is that this does not affect my daily business'' [P277;R45]).
    \item The proposal is too specific and thus difficult to generalize (``This only applies to a small (significant) sector of BAs'' [P394;R17]).
\end{itemize}
Note that many responses provide less detail than the ones listed above and express a short negative justification, e.g., ``It does not make sense for me'') [P127;R120] or ``Not interesting for my working environment'' [P380;R66].

\subsection{RQ4 Which research problems should the RE research community address according to the practitioners?}
\label{sec:Results-RQ4}
After discarding four invalid/empty responses, we split the 99 remaining (64.3\%) into 173 statements.
We removed 16 statements that were comments and not actual suggestions, resulting in 157 suggestions.

We considered using the topic taxonomy from RQ2.4 to classify these suggestions. 
However, we decided against this option because we did not want to force the suggestions into the rigid pre-defined research topic structure. 
As expected, the analysis confirmed some misalignment between the existing research topics and the provided suggestions. 

Table~\ref{tab:TopicSuggestion} presents the 45 codes that emerged from the coding process, grouped into seven categories:
\begin{itemize}
    \item \textit{Process activities}. Similar to the paper topics, these statements focus on specific RE process-related activities. Some codes appear both here and in the topics category (RQ2.4), while some emerged only here (e.g., system scoping), and some appear only in the paper topics (e.g., decision-making).
    \item \textit{Links}. These statements correspond to the need to explore the links among RE and other software engineering activities or artifacts, or even links beyond software engineering.
    \item \textit{Phase}. These statements focus on research about one particular RE phase. We used a slightly finer granularity than in RQ2.4, due to our inductive coding approach. 
    \item \textit{RE and people}. These statements are highly related to human aspects of RE, like the consideration of RE in the organization or putting the human in the loop.
    \item \textit{RE in practice}. These statements are directly related to practical aspects as RE at scale and efficiency of the RE process.
    \item \textit{Specialized}. These statements are highly related to RE, like natural language requirements or data-driven RE.
    \item \textit{Transversal}. These statements refer to larger topic areas not directly dependent on RE, such as education, standards, or tools.
\end{itemize}

\begin{table}
    \centering
    \caption{Codes for research topics suggested}
    \label{tab:TopicSuggestion}
    \begin{tabularx}{\columnwidth}{@{}lXr@{}}
    \toprule
    Category & Codes &Occurrences\\
    \midrule
    Phase & Elicitation~(19), Documentation~(13), Validation~(8), Analysis~(6), Management~(4), Prioritization~(2), Negotiation~(1), Verification~(1)& 54\\[4em]
    Process activities& Traceability~(9), Automation~(8), Evolution~(8), Derivation~(6), Reuse~(4), Modeling~(3), Domain analysis~(3), Formal reasoning~(3), System scoping~(3), Visualization~(2) & 47\\[5em]
    Specialized topics & Requirements quality criteria~(8), RE fundamentals~(7), NL requirements~(7), NFRs~(7), Context-aware RE~(7), Formal requirements~(5), Data-driven RE~(4) & 45\\[5em]
    Transversal areas & RE process~(11), General aspects of RE~(10), RE tools~(6), RE and standards~(4), RE education~(1) & 32\\[3em]
    RE and people & RE in the organization~(12), Human factors~(9), Human in the loop~(6), RE in teams~(3), RE roles~(2)& 32\\[3em]
    RE in practice& RE efficiency~(10), Cost impact~(4), RE at scale~(3), RE effectiveness~(1)& 18\\[2em]
    Links & RE and organization~(7), RE and code~(3), RE and architecture~(2), RE and testing~(2), RE and technology~(1), RE outside CS~(1)& 16\\[3em]
    \bottomrule
    \end{tabularx}
\end{table}

We identified 244 instances in the 157 suggestions (see Table~\ref{tab:TopicSuggestion}). The largest share of statements (94, i.e., 59.9\%) had only one code.
The most codes in one statement was five, (for example: ``Consistent [Requirement properties] representation [Documentation] form: Selection of an appropriate Pattern [Reuse] for NL [NL requirements] or an appropriate modeling form [Formal requirements]'' [R112]). 
There are three clusters of statements: the most suggested categories (Phase, Process activities, and Specialized topics), medium categories (Transversal areas and RE and People), and less suggested categories (RE in practice and Links). 

It is also worth noting that some codes emerged from the analysis of the suggestions. 
For example, some respondents explicitly mentioned the importance of the \textit{Elicitation phase} (``Review effectiveness of Agile methods as compared with traditional methods in the accuracy of requirements elicitation'' [R101]) while others mentioned it more implicitly  (``how to conduct an interview with a customer'' [R43]).

Two last remarks on the information gathered:
\begin{itemize}
    \item The respondents put 19 suggestions (12.2\%) explicitly into context, the majority of them (14) referring to suggestions in agile development, e.g., ``How to convince practitioners that agile does not imply the absence of requirements'' [R76]. None of the other contexts (e.g., ``how to capture and represent ideas regarding innovation projects'' [R121]) was recurrent.
    \item Furthermore, 13 suggestions (8.3\%) contained some type of implicit or explicit general criticism to researchers, e.g., ``Get researcher(s) embedded with real-life system engineering teams and uncover new research areas'' [R44]. We provide more details in Section~\ref{sec:analysis:rq1}.
\end{itemize}

\section{Analysis}\label{sec:analysis}
In this section, we provide our interpretation of the most relevant results reported in the previous section.

\subsection{RE research perception (RQ1)}
\label{sec:analysis:rq1}

In general, the results for RQ1 (Section~\ref{sec:Results-RQ1}) show the majority of the respondents give favourable ratings for the papers. 
However, we would have liked to have seen more \textit{Essential} ratings for papers in our chosen venues. 
These are the conferences that shall lead RE research
forward in the times where agile values play an increasingly vital role and the importance of requirements may become increasingly less evident.

The declining interest over the years is worthy of comment. 
This decline could be due to the overspecialization of papers, which are more focused on challenging research topics rather than on solving real challenges in society and industry. 
Conversely, the decline may simply point out an adoption gap, suggesting that it takes time for industry to adopt research results. 

When examining the effect of participant's role we see some differences. 
On the more critical side, \textit{coaches} find many of the papers \textit{Unimportant} or \textit{Unwise}. 
If one assumes that coaches, in particular, should know what works and does not work in an industrial setting, then their negative perception is particularly concerning.
In addition, the respondents who work with or are influenced by real requirements are less favorable towards RE research.
These include: \emph{designers}, who are directly dependent on the requirements; \emph{consultants}, who view requirements as an essential part of contracts; \emph{managers}, who need to have an overall overview the product; \emph{requirement engineers} and \emph{business analysts}, who produce requirements; \emph{architects}, who, like designers, depend on excellent requirements and sometimes write their own requirements. 
Negative ratings from these participants could result from the perception that much of the published RE work does not relate directly to their biggest practical challenges.

Conversely, ratings from industrial \textit{researchers} are more positive. 
We expect researchers are most familiar with the RE literature, but not necessarily with how the research results apply in day-to-day work, which might explain their more positive views.

Similarly, we notice very positive views from respondents in roles that have some distance from day-to-day RE work, such as \emph{process designer} (highest EW-score) and \emph{multiple roles} (highest E-score).
An explanation for this result might be the huge breadth of RE research.
Only people in roles that have a broad overview of the company can appreciate a large range of topics, while those that have more specific roles have a higher chance of encountering RE research that does not relate to their particular area.

Finally, \emph{testers} and \emph{test managers} stick out as groups with the most favorable perception of RE research.  A possible explanation is that these testers would benefit from the value new practices would bring, without being subjected to any additional workload. 
In addition, they might agree most with the need to improve or change RE in practice.

\subsection{Factors influencing perception (RQ2)}

In this section we overview other factors that influence a respondent's perception of the relevance of RE research.

\subsubsection{Venues}
The slight tendency that papers from RE-specific conferences (REFSQ and IEEE RE) have a higher perceived relevance could result from RE researchers sending their strongest work to these venues because they are the annual meeting points for the RE research community. Furthermore, the reviews from these conferences are usually insightful and provide valuable feedback to authors because the program committee consists of RE experts. 
Conversely, given their broader focus across software engineering, RE papers that appear in non-RE venues may not receive the same level of feedback and may have greater chances of assignment to reviewers less knowledgeable in the paper's research topic.

Even so, the differences are small. 
The number of \textit{Essential} ratings only decreases by a large amount for ESEM. 
This decline may be because methodology is often as important as the results for ESEM papers. 
Therefore, because ESEM papers will typically be more guarded in their claims, the summaries might not be as appealing to respondents.

\subsubsection{Ties to industry}
Based on the percentage of papers with links to industry and the overall perception of relevance, the influence of industrial ties is not clear. 
For instance, Table~\ref{tab:dataset} shows a higher percentage of industry-related papers in IEEE RE than in REFSQ, however respondents rated REFSQ papers as more relevant. 
Therefore, we are not able to draw strong conclusions on the impact of ties to industry. 

Going into more detail, it appears the conference track where the paper appears may influence perception. 
Respondents rated industry track papers slightly more positive than research track papers. 
This increase may be because the topics covered in industry tracks are closer to industry needs, which can result from the characteristics of the industry track calls for papers, including: shorter papers, industry-prevalent program committee, different evaluation guidelines (e.g., less emphasis on research method) and an overall emphasis on practical problems rather than research issues.

An interesting observation is that respondents rated papers with industrial authors (pure industry and mixed author teams) only slightly higher than papers with only academic authors. 
Of course, industry authors submit papers that do not make it through the review process (although we are not able to test this hypothesis). 

\begin{figure}
    \centering
    \includegraphics[width=\columnwidth]{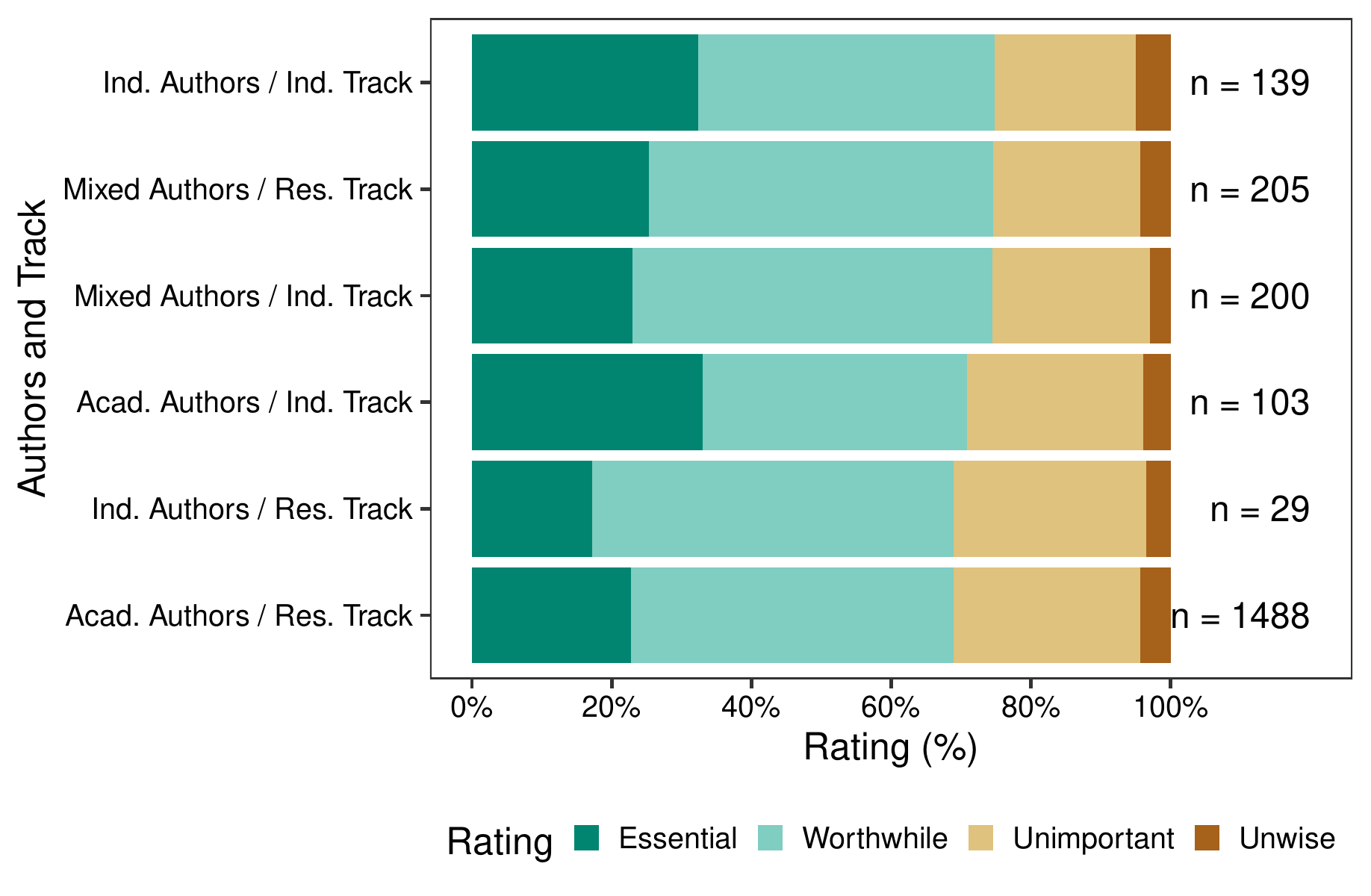}
    \caption{Ratings grouped by author affiliation and submission track}
    \label{fig:RatingsAffiliationTrack}
\end{figure}

To provide more insight, Figure~\ref{fig:RatingsAffiliationTrack} presents the results for the combination of authors and submission tracks.
Interestingly, papers from academic authors appearing in the industry track achieved the highest ratio of \emph{Essential} ratings (E-score: 0.33) while papers from industry authors appearing in the industry track achieved the highest EW-score (0.74). 
Papers from academic authors appearing in the research track had the lowest relevance scores (EW-score: 0.68). 
The majority of the ratings are for this last set of papers.
These 1488 ratings for academic papers in research tracks result in a lower E-score and a lower EW-score than for the 676 ratings for other papers (E-score: 0.23 $<$ 0.27; EW-score: 0.68 $<$ 0.73). 
We can conclude that mixed author teams or submissions to industry tracks are a strong indicator for perceived practical relevance.

\subsubsection{Research methods}

The slight preference respondents have about research method appears to favor abstraction: they consider reference models, methods, and exploratory studies valuable, likely because of their wider claims and perceived applicability appeals to participants. 
Conversely, technology and controlled experiments are likely too narrow to appeal to a broad audience. 
Respondents did not appreciate discussion papers, especially based on the percentage of \textit{Essential} ratings.
In particular, Discussion papers are the most controversial in terms of perception (high U-score). 
A reason for their low score could be the lack of appealing titles and abstracts.

\subsubsection{Research topics}

We found differences in the focus between academia and industry.
For example, goal models and creativity studies are popular RE research areas but are less relevant to the respondents. 
Practitioners likely tend to appreciate the daily challenges they face, such as requirements quality, human aspects, and documentation styles. 

\subsection{Reasons given by practitioners for justifying their perceptions of relevance (RQ3)}
In this section, we separately discuss the reasons for positive perceptions and negative perceptions, based both on the quantitative results presented in Section \ref{sec:Results-RQ3} (Table \ref{tab:PositiveCodes} and Table \ref{tab:NegativeCodes}) and the quotes extracted from the responses (fully available in the replication package).

\subsubsection{Reasons for positive perception}
It is interesting to note that respondents mentioned \textit{problem relevance} and \textit{solution utility} as the reason for positive perceptions with similar frequency, with a slight tendency towards problem relevance. 
In fact, many respondents (41.2\%) gave both reasons as part of their justification.
This observation supports the intuitive belief that for impactful research, a paper has to have both an interesting problem and a plausible solution. 
Still, the strongest opinions have to be with the importance of the problem, especially when it comes to importance in the industry (``relevant in daily project-life'' [P98;R86], ``It is a problem practitioners struggle with'' [P407;R7]).

Examining these results in more detail reveals the most common reasons, as follows:
\begin{itemize}
    \item 
{\it Universal statements of problem relevance prevail over specific ones}. For the ``universal statements'', we refer to general reasons and reasons that relate to industry as a whole (see Section~\ref{sec:Results-RQ3}). For ``specific ones'', we refer to:
\begin{itemize} \item project-related arguments (``I'm working in agile development environments. This items [the subject of the paper] affects my work significantly.'' [P82;R66]),
\item organizational characteristics (``We work with a group of users that differ a lot from each other, so it is important we define requirements based on persona`s and their differences'' [P272;R108]), and
\item individual motivations (``We learned in the undergrad classes that the learned techniques are used in high-risk software, but I haven't seen how it is used in my day to day work'' [P85;R135]).
\end{itemize}
\item {\it Justifications with a potential impact on current practices prevail over research that is more speculative}. 
In other words, practitioners consider research especially valuable when it proposes solutions that:
\begin{itemize}
    \item do not yet exist in industry (``This problem is not addressed in industry. However, almost everyone faces the problem'' [P31;R117]),
    \item exist but are scarce (``Systems (including vehicles) are becoming more autonomous and adaptive, and there is a lot to be done here'' [P198;R5]), and
    \item exist but are difficult to apply or unsatisfactory (``It's quite difficult to specify unambiguous, de-idealized, and measurable NFRs. We usually spend too much time'' [P304;R90]).
\end{itemize}
\end{itemize}

Some additional observations from the data include:
\begin{itemize}
    \item {\it There is a champion among the respondents for every research topic identified in this study}. Respondents refer to general activities or techniques, such as (``\textit{prioritizing} requirements is one of the most challenging tasks for REs'' [P330;R37]) or (``\textit{Traceability} in safety-critical systems is THE important [thing]'' [R124;P114]), but also to detailed issues such as ``Aligning requirements to regulatory standards is extremely important in many industry'' [P2;R116] and ``dependency between requirements is always crucial'' [P126;R56], and general RE skills (``I think that communication, in general, is a success factor of a project; good communication drives to the right goals'' [P105;R124]).
    \item {\it Respondents mention long-running, open RE debates}. For example, how much RE is enough (``Because of the conflict between complete requirements and fast requirements'' [P82;R81]) or agile vs. traditional approaches (``There is a tension between RE rigour and agile; understanding how this might be addressed in practice is critical for the future of RE'' [P364;R123]).
    \item {\it Respondents still lack satisfactory means to handle well-known topics}. Despite their long history both in academia and industry, respondents still mentioned non-functional (quality) requirements and natural language requirements. Practitioners provided different reasons as ``Non-functional requirements do not always get the attention they deserve'' [P128;R74] or reasons related to a particular type, e.g., ``The assessment of requirements ambiguity to achieve this goal [project success] is critical'' [P183;R42]. 
    \item {\it Research can have a tangible value for practitioners}. Some respondents think research results can improve their knowledge or their position in the company, e.g., ``It [the paper] will give me good reasons for RE that managers will understand'' [P28;R22], ``Understanding the applicability of commonly used techniques in different contexts is always helpful as it sharpens the own perception'' [P44;R102].
    \item {\it The research method is important}. A few respondents identify the research method as the main purpose for their assessment, e.g., ``Case studies are critical'' [P184;R78], ``I think an interview-based study can contribute to understanding about this theme'' [P375;R93].
\end{itemize}

\subsubsection{Reasons for negative perception}
Interestingly, the number of respondents who provided reasons for negative responses dropped by 14.4\% (from 119 positive responses to 97 negative responses). 
In addition, the amount of information respondents provided is even less, with a decline from 204 codes to 112 (i.e., from 1.71 codes per response down to 1.15). 
One explanation for this decline could be it is more difficult to explain why something is not relevant than it is to explain why something is relevant.
For relevant research, respondents can apply the research to their daily practices and ``pain points''. 

Although in negative codes the dichotomy between the RE problem addressed and the solution proposed did not emerge as clearly as in the positive codes, we can say that two high-level categories from Table~\ref{tab:NegativeCodes} (Unnecessary and Overall Critique) are closer to the problem space.
Meanwhile, the other three high-level reasons (Disagreement, Low Quality and Difficult) are closer to the solution space. 
Considering this distinction, the number of code instances for the problems is 42 (37.5\%) and 70 for the solutions (62.5\%), a much larger difference than observed for the positive reasons. 
The emphasis on the solution space for unfavorable reasons is the opposite of the slight preference towards problem relevance in positive reasons. 
Therefore, we can say solutions that appear unsound to practitioners are more frequent than non-existing problems as reasons for irrelevant research. 
Even the case in which a respondent provides reasons related to both a problem and a solution, which is frequent in the positive reasons (41.2\%), is the exception in negative reasons (8.2\%).

We also find a tendency for respondents to provide short, unjustified arguments for considering research as not relevant, for example ``not relevant for my work'' [P222;R97], ``it is not necessary'' [P224;R133], ``relatively uninteresting'' [P138;R6] or simply ``fluffy'' [P183;R146]. 
Conversely, only a few respondents provided very precise reasons for their judgement, questioning some assumption of the full research (e.g., ``Assumes that high-level goals and requirements are hierarchical. 
They are not in practice'' [P31;R7]; ``Quantitative analysis of usability is confusing and misleading'' [P41;R115]), sometimes based on experience (``In my experience, to use cross-references they would not help me to identify conflicting requirements'' [P90;R147]). 
In the extreme case, some respondents provided a general reservation to research endeavors or at least to some research (``As a practitioner, I prefer learning about the answers than learning about the problems we all know we have'' [P89;R79]).

Some additional observations from the data include:
\begin{itemize}
    \item {\it Unnecessary research is the dominant negative reason}. A recurring response relates to misalignment with industry practices, e.g. the adoption of formal specifications (``The industry has never gone too far from natural language specifications'' [P214;R132]). However, some respondents explicitly state that the research is not needed in their context, either organizational (``not relevant to my company'' [P428;R81]; ``Agile approach, I do not see this as applicable (at the moment) in the automotive industry'' [P366;R111]) or personal (``probably because of my own attitude'' [P375;R70]).
    \item {\it There are several reasons for criticizing solution utility}. An example of recurrent underlying reason respondents give is: ``interesting, but sounds too theoretical for real project life'' [P396;R92]. Similarly, some respondents express concerns about the efficiency of the proposed methods (``This seems overkill and the expected outcome does not appear to be justifying the efforts'' [P386;R32]). Respondents also complain about excessively narrow focus and give concrete advice to formulate more general approaches (``spend your efforts on helping more BAs!'' [P394;R17]).
    \item {\it The chosen research method is a reason for criticism}. Respondents referred to case studies (``it is limited to a company, if it were a survey with many companies it would have more validity'' [P21;R37]), baseline data (``using a small dataset does not sound interesting'' [P40;R135] and remarkably (4 respondents; 4.1\%) the use of students for experimentation (``Students do not yet have the skills to contribute meaningfully'' [P242;R71]).
    \item {\it Some respondents criticized research others considered positive}. A remarkable example is non-functional requirements, which several respondents identified as important (see the previous section) but others indicated it was not deserving of research because ``NFRs are not that important after all and current prose-based methods are sufficient'' [P304;R41]. Other examples exist, e.g., (``\textit{Traceability} is not a high-priority activity in Requirements Engineering'' [P358;R42]).
\end{itemize}

\subsection{Research problems that the RE research community should address (RQ4)}
We received many suggestions from respondents. 
Some respondents even provided long explanations and motivation, e.g., ``Integration of your techniques and models with Agile development methodologies and Agile architecture methods and tools. In the real world, teams are organized in small agile teams orchestrated by architecture work. At both levels, requirements are very relevant, but the tools used to specify and validate them IMHO are very different from what classical academic research on RE is concerned with. Therefore, analyse those methods and tools and try to provide solutions that are realistic to their problems; as I said, in agile development and agile architecture methods and tools'' [R120].

Conversely, we received 13 suggestions (8.3\%) that expressed fundamental criticism of research, including:
\begin{itemize} 
    \item {\it Skepticism to researchers attitude} (``I think that the community is too focused in producing short term results for scientific publications'' [R5].
    \item {\it Weak evidence} (``do research with examples of reasonable size; solutions that work only on toy example size are of no benefit'' [R103]). 
    \item {\it Unaware of context} (``Check, that the needed prerequisites [of the proposed research] have a certain likelihood to be fulfilled'' [R103]; ``They should first understand the context of my work area'' [R150]).
    \item {\it Lack of applicability of research results} (``Focus within Software Engineering (Requirements Engineering) is on a level that is far too high and abstract (frameworks, philosophy, theory). The industry needs a toolbox with techniques and practices that work'' [R43]).
\end{itemize}

We also want to comment individually on some categories identified in Section~\ref{sec:Results-RQ3}:
\begin{itemize}
    \item {\it Respondents referred to the elicitation and documentation phases most often}. They seldom mentioned other phases like prioritization, negotiation, or verification. While some respondents gave very generic suggestions (e.g., “Techniques for identifying unstated needs”), others were more precise and mentioned topics as diverse as: elicitation by interviews, dealing with tacit knowledge, or how much documentation is enough. For more details, we refer to our online dataset \cite{dataset}.
    \item {\it Respondents also often mentioned the link between RE and organization}. Given that RE lies in the crossroads of several disciplines, it is not surprising to find links to some of them. In particular, we found some responses stressing the link with organizational aspects. Responses may refer to general aspects of this link, i.e. in the Link category (``Visualising requirements and business goals and their interconnections'' [R3]) or emphasize the human factors of such connection (``Transition from research to practice and the associated cost (overcoming management resistance)'' [R44]). Respondents do not seek links to other SE activities like architecture or testing. This observation aligns with the very nature of RE, a discipline that needs to be close to the organizational level.
    \item {\it Three process activities excel over the others: traceability, evolution, and automation}. While these may not be surprising, on the converse, respondents did not want topics like modeling and formal reasoning or more specific activities like visualization and system scoping.
    \item {\it Respondents most often mentioned specialized topics areas that appeared in the reasons given for highly-rated papers}. Specific examples include natural language requirements, non-functional requirements, and the RE process itself (``Transitioning from classic development to agile development while maintaining an existing requirement base'' [R128]). Conversely, the respondents did not mention education frequently, although as one respondent stated: ``RE should be teached [...] may be by practitioners'' [R23].
\end{itemize}

\section{Threats to Validity}
All empirical studies suffer from known and unknown threats to validity. 
Some threats affect the research protocol and the investigated domain. 
By conducting this study collaboratively with authors who have conducted, similar studies, we mitigate threats about the research protocol and the problem domain of RE. 
Wohlin et al.~\cite{Wohlin12}, however, point out other threats that deserve further discussion.

\subsection{Internal Validity}
{\it The context of RE}. We contextualized RE as a subdiscipline of SE. However, it is true that RE is multi-faceted and is present in other contexts such as systems engineering and business process management. We attempted not to impose an SE-only viewpoint on our analysis and discussion, for instance, collecting a significant number of responses from practitioners in roles that are closer to business than to SE like manager or consultant (see Figure~\ref{fig:PerceptionRole}). 

{\it Use of papers}. We used papers as the primary source of information about RE research. Considering that we chose papers from well-respected peer-reviewed conferences in RE, the papers are a reasonable representation of RE research. However, we could have adopted other strategies for answering the research questions, for instance analyzing the practical adoption of research outcomes in industry. A fundamental difference among our approach and this alternative is that in our research design, we could ask a practitioner about an RE topic for which she has no practical experience, while in the alternative approach, a practitioner would have responded according to the RE-related methods she is effectively using in her daily work. Whether this approach would have produced similar results remains an open question.

{\it Practitioners’ perception of research proposals}. 
It was challenging and highly demanding to create summaries. 
We formulated some general guidelines and shared the work among the team. 
However, because the result was not entirely satisfactory despite an internal pilot phase, we conducted another iteration where a subset of authors was in charge of harmonizing the summaries. 
During this process, we decided to create a summary template as a reference (see Section~\ref{sec:design-dataset}). 

{\it Categorization of practitioners’ opinions}. We consistently organized and interpreted the data collected from the participants. 
To support this process, we executed a pilot, involving some members of the research team.

\subsection{Construct Validity}

{\it The meaning of perception}. 
The focus of our study is on how practitioners perceive the relevance of the information in the published research papers. 
Therefore, after analyzing all the collected primary sources, we produced a set of summaries intended to capture the main focus of research. 
Yet, as stated in our earlier paper, ``we are definitely aware that a relevant problem may not be addressed in a relevant way. 
In fact, we are very much aware that the practical relevance of research can truly be judged only after the fact based on the extent to which the ideas have been adopted or not. 
However, our position is that the results of the study can provide a good first indicator of such impact''~\cite{Franch2017}. 

\subsection{Conclusion Validity}
{\it Robustness of the protocol}. This study reuses and evolves research protocols previously executed and evaluated~\cite{Ameller2020,Mendez17}. 
We also performed some preparation~\cite{Franch2017}, executed a pilot, and included authors from the previous similar studies in this replication. 
However, there is still a threat concerning the topics addressed in the papers because the questionnaire does not specifically address a single topic but overall summaries. 
It is still reasonable to assume the topics primarily drive the answers of our respondents.

\subsection{External Validity}
\label{sec:ExternalValidity}
{\it Representativity of RE research by the chosen venues}. 
As the primary source of information, we rely on papers from five world-leading conferences related to RE and SE. Therefore, we excluded other SE conferences where RE topics may be present (e.g., ASE and XP). We also excluded general conferences outside SE, like MODELS or CAiSE, which could include RE-related papers. To partially mitigate this exclusion, we argue that a subset of papers published in IEEE RE and REFSQ present results related to areas other than SE. We also excluded journal papers, which could describe broader aspects and, thus, could have further strengthened our results. 
The exclusion of journal papers is justified by the timeliness and originality of conference papers, besides reducing the complexity of the final study by avoiding the filtering of extended versions and dealing with an excessive amount of data. Additionally, we follow the pattern of one of the previous studies similar to ours~\cite{Lo15}. However, we are aware that industry-relevant research can be published first and only in journals that can present substantially more evidence and contain refined analyses.

{\it Representativity of RE research by the chosen period of time}
Similarly, repeating the study could yield different results as the perception of relevance is time-dependent.
For example, new topics could have momentum not present during this study (e.g., RE for artificial intelligence). 
Mitigating this threat is beyond our control and the intention is to provide an overall, cross-sectional snapshot that can steer an important discussion.

{\it Representativity of the industry by the respondents}. 
We intended to include opinions from a broad range of industry participants who work with software requirements. 
As mentioned in Section~\ref{sec:demographics}, the participants cover a wide range of demographic characteristics, including different countries, levels of experience, roles, team sizes, and industrial domains. We used several channels to recruit participants to minimize sample biases. However, we cannot claim that our sample is representative in a statistical sense since it is practically impossible to characterize the entire population (all practitioners working with requirements).

\section{Discussion}
Not all research has (or should have) the goal of immediate practical application and the quality of research cannot only be judged from the perspective of practical relevance. 
Yet, because RE can be viewed as an applied research field, practical relevance is important. 
RE research results overall should resonate with practitioners. 
This goal is the main motivation of our work that addressed four research questions:
\begin{itemize}
    \item RQ1: While respondents considered a majority of RE research to be relevant, there was a non-negligible percentage that disagreed, showing that the research community still has a challenge ahead.
    \item RQ2: Of the context factors we explored, two had a significant influence on perception. First, respondents had a better perception of solution proposals and more specifically, papers presenting reference models, and a lower perception of discussion papers. 
    Second, respondents viewed empirical papers involving industrial practitioners as more relevant than other papers, especially those involving students.
    \item RQ3: Respondents justified positive opinions on research because of problem relevance and solution utility. Whereas, their negative opinions ranged from unnecessary research and difficulty to apply the proposed solution, to disagreement with the approach and perceived low quality.
    \item RQ4: Practitioners suggested research topics related to phases (mainly elicitation and documentation), activities (with particular mention to traceability, evolution and automation), and specialized topics (as requirement properties), among others.
\end{itemize}

Comparing our results with those from the ``FSE Paper''~\cite{Lo15} and the ``ESEM Paper''~\cite{Carver16} (Section~\ref{sec:RelWork}), we find similarities and differences. 
For example, the primary reasons for the negative perceptions in the current paper and the FSE Paper related to either \textit{Unnecessary research} or \textit{Difficult to apply} in practice. 
Conversely, respondents in the current survey found papers with links to industry had higher perceived relevance than the others, while the ESEM Paper did not report any significant difference on this factor.

Beyond these facts, Table~\ref{tab:studies-comparison} shows the overall results are quite similar. 
The overall proportion of papers that are Essential or Worthwhile (EW-score) is mostly consistent (this paper: 70.5\%, FSE Paper: 71\%, and ESEM Paper: 66.4\%) and the proportion of papers viewed as Unwise is also consistent at +/- 4\%. 
Therefore, we can state, that the practitioners' perception of research does not seem to depend upon any specific SE topic. 
We can hypothesize that researchers follow similar patterns in their publications, regardless of topic area, and practitioners have similar perceptions. 

\begin{table}
    \centering
    \caption{Relevance Perception in RE, SE, and ESE}
    \label{tab:studies-comparison}
    \begin{tabular}{@{}lrrr@{}}
    \toprule
Rating & RE & SE & ESE \\
    \midrule
Essential &24.2\%&20.0\%&17.6\%\\    
Worthwhile &46.3\%&51.0\%&48.8\%\\
Unimportant &25.3\%&25.6\%&29.2\%\\
Unwise &4.2\%&3.4\%&4.6\%\\
    \bottomrule
    \end{tabular}
\end{table}

Several authors have studied research relevance in the last two decades. 
If we focus on topics more related to RE, i.e. software engineering (SE) and information systems (IS), we find several papers on the study of relevance. 
The great majority are based on literature reviews. 
In SE, Garousi et al. recently published a multi-vocal literature review reporting 53 documents on the topic~\cite{Garousi20}. 
The three root causes of low relevance reported in that paper are also identified in our work: wrong assumptions of researchers, lack of connection with industry and wrong identification of research problems. 
In addition, we reported other threats to relevance (answer to RQ3) and identified topics specific to RE. 

\section{Conclusions}
Research relevance has been studied for more than 50 years. 
Practice and research have been qualified as ``poles apart"~\cite{Ryan77}.
Managers have been speculated to be from Mars while academics from Venus~\cite{Baldridge04}. 
Hoping these statements are exaggerations, it is important to reflect on the practical relevance of applied research disciplines, like RE. 
This reflection is the purpose of this paper, a survey-based study involving 153 practitioners worldwide who rated research works published in scientific papers in a way that allowed us provide an answer to four research questions, as summarized in the previous section. 
We argue that the results are valuable for both: {\it researchers}, who learn about factors that can make their research more relevant, and {\it practitioners}, who obtain an overall landscape of the RE state of the art, and {\it educators}.

We would like to see replications of this study following several directions. 
First, an update with research from 2017 onwards, which could uncover new trends about RE research relevance. 
Second, answering the same research questions with different research approaches, e.g. in-depth interviews with practitioners or in-field analysis of research adoption in practice. 
Third, similar studies in other areas like testing or software architecture. 
Aggregation of results would help to validate the outcomes of this study, and could eventually drive possible changes of attitude from practitioners. 
Overall, we hope our research contributes to a culture where academic researchers increasingly present their results in a way that supports uptake in industry.

\section{Acknowledgment}
The authors wish to thank all participating survey respondents, as well as C. Coupette for her support during the elaboration of initial analysis scripts and first interpretations in earlier versions, and O. Dieste for his participation in the elaboration of paper abstracts. X. Franch’s work was partially supported by the Spanish project GENESIS TIN2016-79269-R. D. Mendez' work was partially supported by the KKS foundation through the S.E.R.T. Research Profile project at Blekinge Institute of Technology. G.H. Travassos' work is partically supported by CNPq (grant 304234/2018-4).

	\bibliographystyle{abbrv} 
	\bibliography{re-pract}	
\end{document}